\documentclass[prd,showpacs,preprintnumbers]{revtex4} 

\usepackage{graphicx}
\usepackage{dcolumn}
\usepackage{bm}
\usepackage{latexsym}
\usepackage{amsfonts}
\usepackage{amssymb}
\usepackage{amsmath}
\usepackage{afterpage}

\begin{document}


\title{
Slow-roll Inflation with the Gauss-Bonnet and 
Chern-Simons Corrections
}

\author{Masaki Satoh}
\email{satoh@tap.scphys.kyoto-u.ac.jp}
\affiliation{
 $^{\ast}$Department of Physics,  Kyoto University, Kyoto 606-8501, Japan
}

\date{\today}

\begin{abstract}
We study slow-roll inflation with the Gauss-Bonnet and 
Chern-Simons corrections.
We obtain general formulas 
for the observables: spectral indices,
tensor-to-scalar ratio and circular polarization of 
gravitational waves.
The Gauss-Bonnet term violates the consistency relation
$r=-8n_{\rm T}$.
Particularly, blue spectrum $n_{\rm T}>0$ and scale
invariant spectrum $|8n_{\rm T}|/r\ll1$ of tensor modes
are possible.
These cases require
the Gauss-Bonnet coupling function of 
$\xi_{,\phi}\sim 10^8/M_{\rm Pl}$.
We use examples to show new-inflation-type potential with
$10M_{\rm Pl}$ symmetry breaking scale 
and potential with flat region in 
$\phi\gtrsim10M_{\rm Pl}$
lead to
observationally consistent
blue and scale invariant spectra, respectively. 
Hence, these interesting cases can actually be realized.
The Chern-Simons term produce circularly polarized 
tensor modes.
We show an observation of these signals supports existence of
the Chern-Simons
coupling function of $\omega_{,\phi}\sim 10^8/M_{\rm Pl}$.
Thus, with future observations, we can fix or constrain 
the value of these coupling functions, at the CMB scale.
\end{abstract}

\pacs{}
\maketitle

\section{Introduction}
Currently, the concept of slow-roll inflation is widely accepted
by cosmologists, due to  many precise observations, such
as the 
Wilkinson Microwave Anisotropy Probe (WMAP)\cite{komatsu10}.
We can explain all observations simply, using 
single field slow-roll inflation.
However, future observations might
detect non-standard signals, which will be evidence
for a non-standard inflationary
theory.
The energy scale of inflation is considered to be very high 
compared to current experimental scale.
Thus, we can say that a 
new inflationary theory might contain 
information with regard to high energy
 physics, such as a superstring theory.
Therefore it is important to study  theories beyond
standard single field slow-roll inflation.

One of candidates for a modified
inflationary model
is inflation with the Gauss-Bonnet and Chern-Simons
corrections. These  are the only two
meaningful combinations of second order
curvature terms, as a low energy effective 
theory\cite{weinberg08},
and can also be derived from some superstring 
models\cite{Antoniadis:1993jc}.
Cosmology with the Gauss-Bonnet term can affect 
background evolution and result in a non-singular 
cosmological 
model\cite{Antoniadis:1993jc,Kawai:1998bn,Kawai:1999xn,Kawai:1999pw,Hwang:1999gf,Cartier:2001is,Kawai:1998ab,Soda:1998tr,Kawai:1997mf,Gasperini:1997up,Cartier:2001gc,Leith:2007bu,Guo07,Vazquez:2008wb}
or other interesting 
solutions\cite{Nojiri:2005vv,neupane06a,koivisto07a,koivisto07b,Neupane08,neupane06b,defelice09a,defelice09b,defelice10a,defelice10b}.
In inflation with the Chern-Simons term, primordial 
gravitational waves are circularly 
polarized\cite{Lue:1998mq,Choi:1999zy,Alexander:2004wk,Lyth:2005jf,satoh08a},
which is considered as an interesting target for some
 future observations\cite{Saito:2007kt,Seto:2006hf,taruya}.
Many authors studied Gauss-Bonnet and 
Chern-Simons modified
inflation.
However, {\it slow-roll} inflation with the Gauss-Bonnet terms
is not much studied, except in the 
papers\cite{satoh08b,guo10}.
Taking current observations into consideration,
slow-roll inflation is the most plausible paradigm for inflation.
Therefore 
investigating slow-roll inflation with the Gauss-Bonnet
and Chern-Simons corrections is an important subject.

We have studied this topic in the previous paper\cite{satoh08b}. 
In this paper, we re-formalize slow-roll inflation 
in gravity
with the Gauss-Bonnet and Chern-Simons corrections.
We derive formulas for the observables, namely
the scalar spectral index $n_{\rm S}$, 
the tensor spectral index 
$n_{\rm T}$, 
the tensor-to-scalar ratio $r$ and 
the circular polarization
ratio $\Pi$, in our inflationary model.
The Gauss-Bonnet term violates the
consistency relation $r=-8n_{\rm T}$.
We show if this violation is observationally confirmed, 
the derivative of
the Gauss-Bonnet coupling function at the 
observation scale, namely the Cosmic Microwave Background
(CMB) scale, is fixed.
We study the two typical cases, blue and scale
invariant spectra of tensor modes.
Because blue and scale invariant mean 
$n_{\rm T}>0$ and $|8n_{\rm T}|/r\ll1$, respectively, 
these cases strongly violate this consistency relation.
If these are confirmed in future observations, the Gauss-Bonnet
coupling function must be 
$\xi_{,\phi}\sim 10^8/M_{\rm Pl}$,
at the CMB scale.
Of course, even if no such effects are observed in future, 
we can get the constraint $\xi_{,\phi}\lesssim10^8/M_{\rm Pl}$, 
at least.
However, it is unclear whether these interesting cases 
are consistent with current observations.
Therefore, we perform
calculations in some concrete examples.
We show new-inflation-type potential with symmetry breaking
scale of $10M_{\rm Pl}$ leads to an observationally
consistent blue spectrum;
potential with flat region in $\phi\gtrsim 10M_{\rm Pl}$
leads to a consistent scale invariant spectrum.
In both cases, the almost constant 
$\xi_{,\phi}$ is required.
The Chern-Simons term leads to circular polarization
of gravitational waves.
We show if this circular polarization is detected, we can
fix the derivative of the Chern-Simons coupling function,
at the CMB scale.
It might be $\omega_{,\phi}\sim 10^8/M_{\rm Pl}$.
Therefore, the value of
the Gauss-Bonnet and  Chern-Simons coupling functions at 
the CMB scale 
will be fixed, or constrained, with future observations.

We organize this paper as follows: In section II,
the action we considered is shown and slow-roll inflation
in this action is studied, calculations of
perturbations in slow-roll inflation is in section III,
interesting features of our model are
 mentioned in section IV, 
and section V is for conclusion.

\section{Slow-roll inflation}
\label{sec_slow-roll}
We consider the following action:
\begin{align}
 &S = \int {\rm d}^4 x \sqrt{-g}
 \left[
 \frac{M_{\rm Pl}^2}{2} R -
 \frac{1}{2} \nabla_\alpha \phi \nabla^\alpha \phi -
 V(\phi)
 \right] 
 - \frac{1}{16} \int {\rm d}^4 x \sqrt{-g} \xi(\phi) 
 R_{\rm GB}^2
 + \frac{1}{16} \int {\rm d}^4 x \sqrt{-g} \omega(\phi)
 R \tilde{R}
 \ ,
 \label{S}
\end{align}
where $M_{\rm Pl}^2 \equiv 1/8\pi G$ is the reduced Planck mass,
$R^2_{\rm GB}$ and $R \tilde{R}$ are the Gauss-Bonnet and 
Chern-Simons combinations.
These combinations are defined by
\begin{align}
 R^2_{\rm GB} &\equiv
 R_{\alpha \beta \gamma \delta} R^{\alpha \beta \gamma \delta}
 - 4 R_{\alpha \beta} R^{\alpha \beta} + R^2 \\
 R \tilde{R} &\equiv
 \frac{1}{2} \epsilon^{\alpha \beta \gamma \delta}
 R_{\alpha \beta \rho \sigma} {R_{\gamma \delta}}^{\rho \sigma}
 \ ,
\end{align}
where $\epsilon^{\alpha \beta \gamma \delta}$ is 
the Levi-Civita anti-symmetric tensor.
Hence, the Chern-Simons correction
represents parity violation in gravity.
This action is composed of the 
Einstein-Hilbert term and a canonical
scalar field which couples to the Gauss-Bonnet and Chern-Simons
terms through the coupling functions $\xi(\phi)$ and 
$\omega(\phi)$. 
Note that only these two types of higher curvature corrections,
namely the Gauss-Bonnet and Chern-Simons terms, 
are essential as a low energy effective theory\cite{weinberg08}.
Here, we take these terms as small corrections.
In principle, coupling functions must be fixed
by quantum gravity, such as
a superstring theory. However, it is impossible to determine
these coupling functions, because we haven't succeeded to construct
quantum gravity. Therefore, we treat these as 
free functions and attempt to fix or 
constrain these coupling functions from
cosmological observations. Through these attempts, 
we may get new information with regard to quantum
gravity, because these 
coupling functions are related to properties
of quantum gravity.

Here, as a background metric, we choose the flat 
Friedmann-Robertson-Walker(FRW) metric:
\begin{align}
 {\rm d}^2 s = - {\rm d} t^2 + a^2(t) \delta_{ij} 
 {\rm d} x^i {\rm d} x^j \ , \label{frw}
\end{align}
where $a(t)$ denotes a scale factor which describes cosmic
expansion. Variating Eq.(\ref{S}) in the FRW
metric, Eq.(\ref{frw}),
we get background equations:
\begin{align}
 &3 M_{\rm Pl}^2 H^2 = 
 \frac{1}{2} \dot{\phi}^2 + V + \frac{3}{2} H^3 \dot{\xi}
 \label{friedmann0}
 \\
 &\ddot{\phi} + 3 H \dot{\phi} + \frac{3}{2} H^2
 (\dot{H} + H^2) \xi_{,\phi} + V_{,\phi} = 0 \ ,
 \label{eom0}
\end{align}
where a dot denotes a 
time derivative and the Hubble parameter $H$
is defined by $H \equiv \dot{a} / a$. 
Because of parity 
symmetry of Eq.(\ref{frw}), the Chern-Simons correction
cannot affect these background equations.
We take $\dot{\phi}<0$ throughout this paper.

Here we focus on the case in which scalar field is 
slowly rolling and its friction term is dominating
Eq.(\ref{eom0}).
In other words,  we take
slow-roll approximations:
\begin{align}
 \frac{\dot{\phi}^2}{M_{\rm Pl}^2 H^2} \ll 1\ , \quad
 \frac{\ddot{\phi}}{H \dot{\phi}} \ll 1
 \ .
 \label{slow-roll}
\end{align}
In this paper, we consider the Gauss-Bonnet term as 
a {\it small} correction. Hence, it must be 
much smaller than the Einstein-Hilbert term, 
energetically
speaking.
The following inequality, hence, must be satisfied:
\begin{align}
 \frac{H^3 \dot{\xi}}{2 M_{\rm Pl}^2 H^2} 
 = \frac{H \dot{\xi}}{2 M_{\rm Pl} ^2}
 \ll 1 
 \ . \label{gb_is_small}
\end{align}
Under Eqs.(\ref{slow-roll}) and (\ref{gb_is_small}),
the background equation, Eq.(\ref{friedmann0}), 
become
\begin{align}
 H^2 = \frac{V}{3 M_{\rm Pl}^2}\ .
 \label{friedmann_slow}
\end{align}
This equation doesn't contain the Gauss-Bonnet correction
since it is energetically negligible, of course. 
The time derivative of the Hubble parameter becomes
\begin{align}
 \frac{\dot{H}}{H^2} = \frac{({H^2}\dot{)}}{2 H^3}
 = \frac{V_{,\phi} \dot{\phi}}{6M_{\rm Pl}^2 H^3}
 = \frac{1}{2} \frac{V_{,\phi}}{V} \frac{\dot{\phi}}{H}
 \ .
 \label{dotH_slow}
\end{align}
The another background equation, 
Eq.(\ref{eom0}), is
\begin{align}
 \dot{\phi} = 
 - \frac{V_{,\phi}}{3 H}
 - \frac{1}{2} H^3 \xi_{,\phi} 
 - \frac{1}{2} H \dot{H} \xi_{,\phi}
 \ .
\end{align}
The last term in the right-hand side can be written as
\begin{align}
 - \frac{1}{2} H \dot{H} \xi_{,\phi}
 = - \frac{1}{4} \frac{V_{,\phi}}{V} H^2 \dot{\xi}
 = - \frac{V_{,\phi}}{3 H} \frac{H \dot{\xi}}{4 M_{\rm Pl}^2}
 \ .
 \label{last_term}
\end{align}
Hence, this term is negligible
compared with the first term in the right-hand side
of
Eq.(\ref{gb_is_small}),
and Eq.(\ref{eom0})
becomes
\begin{align}
 \frac{\dot{\phi}}{M_{\rm Pl} H} = 
 - M_{\rm Pl} \frac{V_{,\phi}}{V}
 - \frac{V \xi_{,\phi}}{6 M_{\rm Pl}^3} \ .
 \label{eom_slow}
\end{align}
This equation shows that the Gauss-Bonnet coupling function $\xi$ works
as effective potential.
In our inflationary model, slow-roll equations are 
Eq.(\ref{friedmann_slow}) and (\ref{eom_slow}). We can see
only the equation of motion, Eq.(\ref{eom_slow}), is modified.
Here, we define the new function $\zeta$:
\begin{align}
 \zeta(\phi) \equiv V_{,\phi}
 + \frac{V^2 \xi_{,\phi}}{6 M_{\rm Pl}^4} \ . 
 \label{def_zeta}
\end{align}
Eq.(\ref{eom_slow}) becomes
\begin{align}
 \frac{\dot{\phi}}{M_{\rm Pl} H}  =
 - \frac{M_{\rm Pl}}{V} \zeta \ .
 \label{dphizeta}
\end{align}
In our model, the 
expression for the e-folding number $N$ is written as
\begin{align}
 N = \int {\rm d}t H
 = \int^0_\phi {\rm d}\phi \frac{H}{\dot{\phi}}
 = \frac{1}{M_{\rm Pl}^2}
 \int^\phi_0 {\rm d}\phi  \frac{V}{\zeta} 
 = \frac{1}{M_{\rm Pl}^2}
 \int^\phi_0 {\rm d}\phi
 \frac{1}{V_{,\phi}/V+V\xi_{,\phi}/6M_{\rm Pl}^4}
 \ ,
 \label{e-fold_0}
\end{align}
where we assume that potential has a minimum at $\phi = 0$.

We must consider consistency conditions for 
Eqs.(\ref{friedmann_slow}) and (\ref{dphizeta}).
First, we check slow-roll conditions,
Eqs.(\ref{slow-roll}):
\begin{align}
 \frac{\dot{\phi}^2}{M_{\rm Pl}^2 H^2}
 &= M_{\rm Pl}^2 \frac{\zeta^2}{V^2} \ll 1 \ ,
 \\
 \frac{\ddot{\phi}}{H \dot{\phi}}
 &= \frac{1}{\dot{\phi}} 
 \frac{\rm d}{{\rm d}t} \frac{\dot{\phi}}{H}
 + \frac{\dot{H}}{H^2}
 = -M_{\rm Pl}^2 \frac{\rm d}{{\rm d}\phi}\frac{\zeta}{V}
 -\frac{M_{\rm Pl}^2}{2}\frac{\zeta V_{,\phi}}{V^2}
 = \frac{M_{\rm Pl}^2}{2} \frac{\zeta V_{,\phi}}{V^2}
 - M_{\rm Pl}^2 \frac{\zeta_{,\phi}}{V}
 \ll 1 \ .
\end{align}
Second, the condition for smallness of 
the Gauss-Bonnet term, Eq.(\ref{gb_is_small}), is
\begin{align}
 \frac{H \dot{\xi}}{2 M_{\rm Pl}^2}
 &= \frac{H\dot{\phi}}{2M_{\rm Pl}^2}\xi_{,\phi}
 = -\frac{1}{2} 
 \frac{H^2}{V} \zeta \xi_{,\phi}
 = -\frac{1}{6M_{\rm Pl}^2} \zeta
 \frac{6M_{\rm Pl}^4}{V^2} 
 \left(\zeta-V_{,\phi}\right)
 = M_{\rm Pl}^2 \frac{\zeta V_{,\phi}}{V^2}
 - M_{\rm Pl}^2 \frac{\zeta^2}{V^2}
 \ll 1
 \ . \label{sigma}
\end{align}
Here, we define three slow-roll parameters:
\begin{align}
 \alpha \equiv 
 \frac{M_{\rm Pl}^2}{2} \frac{\zeta V_{,\phi}}{V^2}
 \ , \quad
 \beta \equiv \frac{M_{\rm Pl}^2}{2}
 \frac{\zeta^2}{V^2} \ , \quad
 \gamma \equiv 
 M_{\rm Pl}^2 \frac{\zeta_{,\phi}}{V}
 \ .
 \label{slow-roll_param}
\end{align}
It is clear that 
if $|\alpha|,\beta,|\gamma| \ll 1$ are realized, all consistency
conditions are satisfied and slow-roll inflation occurs.
In the case of conventional slow-roll inflation,
$\zeta=V_{,\phi}$ is realized.
Hence, 
$\alpha$ and $\beta$ becomes conventional $\epsilon$, and
$\gamma$ becomes $\eta$.
For later convenience, we derive expressions for $\dot{H}/H^2$, 
$\dot{\phi}$ and $H\dot{\xi}/M_{\rm Pl}^2$
in slow-roll parameters. From Eqs.(\ref{dotH_slow}),
(\ref{dphizeta}) and (\ref{sigma}), we get
\begin{align}
 \frac{\dot{H}}{H^2} = - \frac{1}{2} \frac{V_{,\phi}}{V}
 \frac{M_{\rm Pl}^2}{V}\zeta
 = - \frac{M_{\rm Pl}^2}{2} \frac{\zeta V_{,\phi}}{V^2}
 = - \alpha
 \ , \quad
 \frac{\dot{\phi}}{M_{\rm Pl}H} = -\sqrt{2\beta} \ , \quad
 \frac{H\dot{\xi}}{M_{\rm Pl}^2} = 4 (\alpha-\beta) \ .
 \label{dotH_param}
\end{align}
Remember that we are taking $\dot{\phi} < 0$.
As in a conventional scenario, 
the time derivative of the Hubble parameter
is small.
In the case that $\alpha$ can be treated as a constant,
we can solve the equation for $H$ easily:
\begin{align}
 H = \frac{1}{\alpha t} \ , \quad
 a = a_0 |t|^{1/\alpha} \ ,
\end{align}
where $a_0$ is a constant of integration.
Because we choose $H>0$, 
$t>0$ for $\alpha>0$ and $t<0$ for $\alpha<0$.

The most important difference between our model and a 
conventional
inflation is that, in our model super inflation can 
be realized, while in a conventional one it cannot.
In our model, the slow-roll parameter $\alpha$ can take 
a negative value. 
From Eqs.(\ref{dotH_param}), we can see
this leads to super inflation, $\dot{H}>0$.
We can expect that if this super inflation phase corresponds
to the CMB
 scale, an observationally interesting signal
could be detected in
future experiments, and this signal 
might be regarded as evidence for existence of
gravitational higher curvature terms.

Here, we show an 
interesting effect of the Gauss-Bonnet correction
in slow-roll inflation.
Let us consider general potential $V(\phi)$, which does 
not
satisfies conventional
slow-roll conditions.
If we set 
\begin{align}
 \xi(\phi) = \frac{6 M_{\rm Pl}^4}{V(\phi)} \ ,
\end{align}
$\zeta(\phi)=0$ is realized. This makes 
all slow-roll parameters zero, and inflation occurs.
This shows that, even if we use extremely steep
potential, slow-roll inflation can be achieved due to
the Gauss-Bonnet correction.

\section{Perturbations}
In this section, we perform perturbative calculations on
the background of slow-roll inflation.
First, we study scalar and tensor modes of perturbations. 
Note that
no vector modes plays an important role since there is no
source for vectors. 
Second, we derive expressions for 
reconstructing the value of 
potential and coupling functions
using observables.
Finally, we briefly comment on the Lyth bound in our model.

\subsection{Scalar perturbations}
Here, we consider scalar perturbations in the flat FRW
background. The metric is
\begin{align}
 {\rm d} s^2 = a^2 (\tau)
 \left[
 - (1 + 2 A) {\rm d} \tau^2
 + B_{|i} {\rm d} \tau {\rm d} x^i
 + (\delta_{ij} + 2 \psi \delta_{ij} + 2 E_{|ij})
 {\rm d} x^i {\rm d} x^j
 \right] \ ,
\end{align}
where $\tau$ is the conformal time ${\rm d}\tau\equiv{\rm d}t/a$
and a bar denotes a spacial derivative. Latin indices run
from $1$ to $3$.
In the case of slow-roll inflation, $\tau$ becomes
\begin{align}
 \tau = \frac{1}{a_0}
 \frac{1}{1-1/\alpha} t |t|^{-1/\alpha}
 = \frac{t}{1-1/\alpha} \frac{1}{a}
 = - \frac{1}{1-\alpha} \frac{1}{aH} \ ,
\end{align}
where $\tau < 0$.
We fix gauge as $E=\delta\phi=0$, where $\delta\phi$
is a perturbation of scalar field $\phi$.
This gauge is useful in this case, because
the perturbative quantity $\psi$ can be 
treated as a master variable and $-\psi$ coincide with 
the gauge invariant 
comoving curvature perturbation ${\cal R}$.
Of course, this is a complete choice of gauge.
Here, we concentrate on $\psi$, because what we need is the
power spectrum of $\psi$ in super-horizon scale.
Expanding $\psi$ in Fourier modes,
\begin{align}
 \psi = \frac{1}{M_{\rm Pl}}
 \int \frac{{\rm d} k^3}{(2 \pi)^3} \psi_{\bf k}
 {\rm e}^{i {\bf k \cdot x}}\ ,
\end{align}
the action for $\psi$ becomes
\begin{align}
 S = \frac{1}{2} \int {\rm d} \tau
 \int \frac{{\rm d}k^3}{(2 \pi)^3} A_\psi^2
 \left( |\psi'_{\bf k}|^2 - C_\psi^2 k^2 |\psi_{\bf k}|^2 \right)
 \ ,
\end{align}
where a prime denotes a conformal time derivative and
\begin{align}
 A_\psi^2 &\equiv a^2
 \left( \frac{1 - \sigma/2}{1 - 3\sigma/4} \right)^2
 \left(
 \frac{\dot{\phi}^2}{M_{\rm Pl}^2 H^2} 
 + \frac{3}{8} \frac{\sigma^2}{1 - \sigma/2}
 \right)
 \ , \\
 C_\psi^2 &\equiv 1
 + \frac{a^2}{A_\psi^2}
 \left( \frac{\sigma}{1 - 3\sigma/4} \right)^2
 \left[
 - \frac{\sigma}{16} + \frac{\ddot{\xi}}{16 M_{\rm Pl}^2}
 + \frac{\dot{H}}{2 H^2} \left( 1 - \frac{\sigma}{2} \right)
 \right]
 \ ,
\end{align}
and $\sigma$ is defined by
\begin{align}
 \sigma \equiv \frac{H \dot{\xi}}{M_{\rm Pl}^2} \ .
\end{align}
This action contains only $\xi$. This is because 
scalar perturbations in the flat FRW background have
parity symmetry.
Let us define the new variable $\Psi_{\bf k}$, by 
$\Psi_{\bf k}\equiv A_\psi \psi_{\bf k}$, and the above
action becomes
\begin{align}
 S = \frac{1}{2} \int {\rm d} \tau
 \int \frac{{\rm d} k^3}{(2\pi)^3}
 \left(
 |\Psi'_{\bf k}|^2 - C_\psi^2 k^2 |\Psi_{\bf k}|^2
 + \frac{A''_\psi}{A_\psi} |\Psi_{\bf k}|^2
 \right) \ .
\end{align}
This leads to the equation of motion:
\begin{align}
 \Psi''_{\bf k} +
 \left(C_\psi^2 k^2 - \frac{A''_\psi}{A_\psi} \right)
 \Psi_{\bf k} = 0 \ .
 \label{eom_s0}
\end{align}

Here, we impose slow-roll approximation, and derive
the expressions for $A_\psi^2$ and $C_\psi^2$ up to
first order in slow-roll parameters. Remember that
the time derivative of 
the Hubble parameter $H$ is first order.
In computing primordial power spectra, we
treat slow-roll parameters as constants of time.
We get the following equations:
\begin{align}
 \sigma = 4 (\alpha - \beta) \ , \quad
 \frac{\dot{\phi}^2}{M_{\rm Pl}^2 H^2} = 2 \beta \ ,
 \quad
 \frac{\ddot{\xi}}{M_{\rm Pl}^2} =
 \frac{\dot{\sigma}}{H} 
 - \frac{\dot{H}}{H^2} \sigma = 0 \ ,
\end{align}
and
\begin{align}
 A_\psi^2 = a^2 \left( 1 + \sigma / 2 \right) 2 \beta
 = 2 a^2 \beta \ , \quad
 C_\psi^2 = 1 \ .
\end{align}
Let us consider 
the effective potential $A''_\psi/A_\psi$. For this purpose,
we derive some equations:
\begin{align}
 \frac{\dot{\beta}}{H \beta} &= \frac{2 \dot{\zeta}}{H \zeta}
 - \frac{2 \dot{V}}{H V} 
 = \frac{2 \zeta_{,\phi}}{\zeta} \frac{\dot{\phi}}{H} 
 - \frac{2 V_{,\phi}}{V} \frac{\dot{\phi}}{H}
 = - 2M_{\rm Pl}^2\frac{\zeta_{,\phi}}{V}
 + 2M_{\rm Pl}^2 \frac{\zeta V_{,\phi}}{V^2}
 = 4\alpha - 2\gamma \ ,
 \\
 \frac{A'_\psi}{A_\psi} &= \frac{a}{2} 
 \frac{(A_\psi^2\dot{)}}{A_\psi^2}
 = a H + \frac{a}{2} \frac{\dot{\beta}}{\beta}
 = a H (1 + 2 \alpha - \gamma)
 = \frac{1}{-\tau} \frac{1+2\alpha-\gamma}{1-\alpha}
 = \frac{1+3\alpha-\gamma}{-\tau} \ .
\end{align}
Hence
\begin{align}
 \frac{A''_\psi}{A_\psi} = \frac{\rm d}{{\rm d}\tau}
 \left(\frac{A'_\psi}{A_\psi}\right)
 + \left(\frac{A'_\psi}{A_\psi}\right)^2
 = \frac{1+3\alpha-\gamma}{\tau^2}
 + \frac{1+6\alpha-2\gamma}{\tau^2}
 = \frac{2+9\alpha-3\gamma}{\tau^2}
 = \frac{\nu_\psi^2-1/4}{\tau^2} \ ,
\end{align}
where $\nu_\psi$ is defined by
\begin{align}
 \nu_\psi \equiv \frac{3}{2}+3\alpha-\gamma \ .
\end{align}
Eq.(\ref{eom_s0}) becomes
\begin{align}
 \Psi''_{\bf k} +
 \left(k^2 - \frac{\nu_\psi^2-1/4}{\tau^2}\right)
 \Psi_{\bf k} = 0 \ .
 \label{eom_s1}
\end{align}

To quantize the quantity $\Psi$, we replace $\Psi$ with
the operator $\hat{\Psi}$ and expand this as
\begin{align}
 \hat{\Psi_{\bf k}} = v_{\bf k}(\tau) \hat{a}_{\bf k}
 + v_{\bf -k}^*(\tau) \hat{a}_{\bf -k}^\dagger \ ,
\end{align}
where $\hat{a}_{\bf k}$ and $\hat{a}_{\bf -k}^\dagger$
are annihilation and creation operators.
The mode function $v_{\bf k}(\tau)$ obeys the same 
equation as $\Psi_{\bf k}$ does, namely Eq.(\ref{eom_s1}):
\begin{align}
 v''_{\bf k} +
 \left(k^2 - \frac{\nu_\psi^2-1/4}{\tau^2}\right)
 v_{\bf k} = 0 \ .
 \label{eom_s2} 
\end{align}
For solving this equation, we must impose an initial condition.
Because effective potential vanishes
in asymptotic past, we can choose the Bunch-Davies
vacuum at $\tau\rightarrow-\infty$:
\begin{align}
 v_{\bf k} = \frac{1}{\sqrt{2k}}
 {\rm e}^{-i k\tau} \ .
\end{align}
Under this condition, Eq.(\ref{eom_s2}) has the analytic
solution
\begin{align}
 v_{\bf k}(\tau) = \frac{\sqrt{-\pi\tau}}{2}
 {\rm e}^{i(\pi/4 + \pi\nu_\psi/2)}
 {\rm H}^{(1)}_{\nu_\psi} (-k\tau) \ ,
\end{align}
where ${\rm H}^{(1)}_{\nu_\psi}$ is the first kind 
of Hankel function.

Here, we derive the expression for
the power spectrum of scalar perturbation
${\cal P}_{\rm S}$ in super-horizon, which is defined by
\begin{align}
 \langle0| \hat{\psi}^\dagger \hat{\psi} |0\rangle
 = \int{\rm d}(\log k) \frac{1}{M_{\rm Pl}^2}
 \frac{1}{2\pi^2} k^3
 \frac{|v_{\bf k}|^2}{A_\psi^2}
 = \int{\rm d}(\log k) {\cal P}_{\rm S}(k) \ .
\end{align}
In super-horizon, $-k\tau\ll1$, $v_{\bf k}(\tau)$
has the asymptotic form
\begin{align}
 v_{\bf k}(\tau) 
 = \sqrt{\frac{-\tau}{2(-k\tau)^3}} 
 {\rm e}^{i(-\pi/4+\pi\nu_\psi/2)}
 \frac{\Gamma(\nu_\psi)}{\Gamma(3/2)}
 \left(\frac{-k\tau}{2}\right)^{3/2-\nu_\psi} \ .
\end{align}
Hence ${\cal P}_{\rm S}$ becomes
\begin{align}
 {\cal P}_{\rm S}(k) = \frac{1}{M_{\rm Pl}^2} \frac{1}{4\pi^2} 
 \frac{1}{A_\psi^2\tau^2}
 \left(\frac{\Gamma(\nu_\psi)}{\Gamma(3/2)}\right)^2
 \left(\frac{-k\tau}{2}\right)^{3-2\nu_\psi}
 = \frac{1}{8\pi^2\beta} 
 \left(\frac{H}{M_{\rm Pl}}
 \right)^2
 \left(\frac{-k\tau}{2}\right)^{3-2\nu_\psi} \ .
 \label{psex}
\end{align}
The scalar power spectrum is of order 
$H^2/(\beta M_{\rm Pl}^2)$, and this is easily understood
because
 ${\cal P}_{\rm S}\sim (H^2/\dot{\phi})^2\sim H^2/(\beta M_{\rm
Pl}^2)$, in which we use Eq.(\ref{dotH_param}).
The scalar spectral index $n_{\rm S}$ becomes
\begin{align}
 n_{\rm S} - 1 = 3-2\nu_\psi = -6\alpha + 2\gamma \ .
 \label{nsex}
\end{align}
If we take $\xi=0$, 
the slow-roll parameters $\alpha$ and $\gamma$ coincide
with the conventional ones
$\epsilon$ and $\eta$, respectively.
Hence, in this case, the expression for $n_{\rm S}$
takes ordinary form.

\subsection{Tensor perturbations}
In this subsection, we compute perturbations of
tensor modes. The metric is 
\begin{align}
 {\rm d}s^2 = a^2(\tau)
 \left[
 -{\rm d}\tau^2 + (\delta_{ij} + h_{ij}){\rm d}x^i{\rm d}x^j
 \right] \ ,
\end{align}
where $h_{ij}$ denotes a transverse-traceless tensor 
on a constant-time hypersurface. Hence it satisfies
${h_i}^i=h_{ij|i}=0$. 
We expand $h_{ij}$, which is gauge-invariant
because tensor modes have no gauge freedom, in Fourier modes and
circular polarization:
\begin{align}
 h_{ij} = \frac{\sqrt{2}}{M_{\rm Pl}} \sum_{\pm}
 \int \frac{{\rm d}k^3}{(2\pi)^3}
 \phi^\pm_{\bf k} {\rm e}^{i{\bf k\cdot x}}
 p^\pm_{{\bf k},ij} \ ,
\end{align}
where 
$p^\pm_{{\bf k},ij}$ is the polarization tensor for
circular polarization and
$\pm$ denotes the helicity of each polarization mode.
Note that $\sqrt{2}$ and $M_{\rm Pl}$ are
for later convenience.
The action for $\phi^\pm_{\bf k}$ becomes
\begin{align}
 S = \sum_\pm \frac{1}{2} \int {\rm d}\tau
 \frac{{\rm d}k^3}{(2\pi)^3} A_{\rm T}^2
 \left(
 |(\phi^\pm_{\bf k})'|^2 - C_{\rm T}^2
 |\phi^\pm_{\bf k}|^2
 \right) \ ,
 \label{s_t0}
\end{align}
where
\begin{align}
 A_{\rm T}^2 \equiv a^2
 \left(1-\frac{\sigma}{2} \mp \frac{1}{M_{\rm c}}
 \frac{k}{a}\Omega
 \right) \ , \quad
 C_{\rm T}^2 \equiv 1 + \frac{a^2}{A_{\rm T}^2}\frac{\sigma}{2}
 -\frac{\ddot{\xi}}{2M_{\rm Pl}^2} \frac{a^2}{A_{\rm T}^2}
 \ .
 \label{atct}
\end{align}
The mass scale 
$M_{\rm c}$ corresponds to
 the UV cut-off scale of our model and
$\Omega$ is defined by
\begin{align}
 \Omega \equiv \frac{1}{2} \frac{M_{\rm c}}{M_{\rm Pl}}
 \frac{\dot{\omega}}{M_{\rm Pl}} \ .
 \label{def_Omega}
\end{align}
Note that we omit $\pm$ symbols in the expressions
for $A_{\rm T}^2$ and $C_{\rm T}^2$
for simplicity.
The Chern-Simons coupling function $\omega$
doesn't appear in the background and scalar 
perturbations, but, at the same time,
this appears in tensor perturbations,
 because {\it circular} modes clearly violate parity 
symmetry.

From Eqs.(\ref{atct}), we can see that if $|\Omega|>1$, 
$A_{\rm T}^2$ takes a negative value in one of helicity modes
at the cut-off scale $k/a=M_{\rm c}$. This is very important
because it implies the appearance of ghost modes in gravitational
waves.
Since we don't know  
proper treatment for ghost modes, we impose
the condition $|\Omega|<1$. 
For simplicity of 
following calculation, we take $\Omega$ as a constant, 
which is justified
in many models because $\phi$ is slowly rolling.
We can always treat the Chern-Simons correction 
$k\Omega/M_{\rm c}a$ as a small correction, because
$|\Omega|<1$ and $k/a<M_{\rm c}$.
Of course, $|\Omega|>1$ model might lead to very drastic
results and such model is interesting itself. 
However, 
we don't analyze this situation in this paper.

Introducing the new variable 
$\mu_{\bf k}\equiv A_{\rm T} \phi^\pm_{\bf k}$,
Eq.(\ref{s_t0}) can be rewritten as
\begin{align}
 S = \sum_\pm \frac{1}{2} \int {\rm d}\tau
 \int\frac{{\rm d}k^3}{(2\pi)^3}
 \left(
 |\mu'_{\bf k}|^2 - C_{\rm T}^2 k^2 |\mu_{\bf k}|^2
 + \frac{A''_{\rm T}}{A_{\rm T}}|\mu_{\bf k}|^2
 \right) \ .
\end{align}
Hence the equation of motion for $\mu_{\bf k}$ is
\begin{align}
 \mu''_{\bf k} +
 \left(
 C_{\rm T}^2 k^2 - \frac{A''_{\rm T}}{A_{\rm T}}
 \right) \mu_{\bf ki} = 0 \ ,
 \label{eom_t0}
\end{align}
where we also omit $\pm$ in a $\mu_{\bf k}$ expression.

Considering slow-roll inflation, we compute spectra
up to first order in slow-roll parameters and $\Omega$.
The quantities $A_{\rm T}$ and $C_{\rm T}$
become
\begin{align}
 A_{\rm T}^2 = a^2
 \left( 
 1 - 2\alpha + 2\beta \mp \frac{k\Omega}{M_{\rm c}a}
 \right) \ , \quad
 C_{\rm T}^2 = 1 + 2\alpha - 2\beta \ ,
\end{align}
and we get
\begin{align}
 \frac{A'_{\rm T}}{A_{\rm T}}
 = \frac{a}{2} \frac{(A_{\rm T}^2\dot{)}}{A_{\rm T}^2}
 = \frac{a}{2}\left(
 2 H \pm \frac{k\Omega}{M_{\rm c}a}H
 \right)
 = aH \pm k \frac{\Omega}{2M_{\rm c}}H
 = \frac{1+\alpha}{-\tau}
 \pm k \frac{H}{2M_{\rm c}} \Omega \ .
\end{align}
Effective potential becomes
\begin{align}
 \frac{A''_{\rm T}}{A_{\rm T}}
 = \frac{\rm d}{{\rm d}\tau}
 \left(\frac{A'_{\rm T}}{A_{\rm T}}\right)
 + \left(\frac{A'_{\rm T}}{A_{\rm T}}\right)^2
 = \frac{1+\alpha}{\tau^2}
 + \frac{1+2\alpha}{\tau^2}
 \pm \frac{k}{-\tau}\frac{H}{M_{\rm c}} \Omega
 = \frac{2+3\alpha}{\tau^2}
 \pm \frac{k}{-\tau}\frac{H}{M_{\rm c}} \Omega
 = \frac{\nu_{\rm T}^2-1/4}{\tau^2}
 \pm \frac{k}{-\tau}\frac{H}{M_{\rm c}} \Omega 
 \ ,
\end{align}
where $\nu_{\rm T}$ is defined by
\begin{align}
 \nu_{\rm T} \equiv \frac{3}{2} + \alpha \ .
\end{align}
Hence, Eq.(\ref{eom_t0}) becomes
\begin{align}
 \mu''_{\bf k} +
 \left(
 C_{\rm T}^2 k^2
 - \frac{\nu_{\rm T}^2-1/4}{\tau^2}
 \mp \frac{k}{-\tau} \frac{H}{M_{\rm c}} \Omega
 \right) \mu_{\bf k} = 0 \ .
 \label{eom_t1}
\end{align}

Let us quantize perturbations. We promote $\mu_{\bf k}$
to the operator $\hat{\mu}_{\bf k}$ and expand in
the annihilation and creation operators $\hat{a}_{\bf k}$
and $\hat{a}^\dagger_{\bf -k}$:
\begin{align}
 \hat{\mu}_{\bf k}
 = u_{\bf k}(\tau) \hat{a}_{\bf k}
 + u^*_{\bf -k}(\tau) \hat{a}^\dagger_{\bf -k} \ .
\end{align}
The equation for $u_{\bf k}(\tau)$ is the same as that for
$\mu_{\bf k}(\tau)$, Eq.(\ref{eom_t1}):
\begin{align}
 u''_{\bf k} +
 \left(
 C_{\rm T}^2 k^2
 - \frac{\nu_{\rm T}^2-1/4}{\tau^2}
 \mp \frac{k}{-\tau} \frac{H}{M_{\rm c}} \Omega
 \right) u_{\bf k} = 0 \ .
 \label{eom_t2}
\end{align}
As in the case of scalar modes, effective potential vanishes at
asymptotic past. Thus we can use the Bunch-Davies vacuum
as the initial condition for Eq.(\ref{eom_t2}).
We take at $\tau\rightarrow -\infty$:
\begin{align}
 u_{\bf k} = \frac{1}{\sqrt{2C_{\rm T}k}}
 {\rm e}^{-iC_{\rm T}k\tau} \ .
\end{align}
Hence, the solution for Eq.(\ref{eom_t2}) is
\begin{align}
 u_{\bf k} = {\rm e}^{-iC_{\rm T}k\tau}
 (-2C_{\rm T}k\tau)^{\nu_{\rm T}} \sqrt{-\tau}
 {\rm e}^{-i(\pi/4+\pi\nu_{\rm T}/2)}
 {\rm U}\left(
 \frac{1}{2}+\nu_{\rm T}\mp \frac{i}{2C_{\rm T}}
 \frac{H}{M_{\rm c}} \Omega,\ 
 1+2\nu_{\rm T}, \ 
 2 i C_{\rm T} k\tau
 \right) 
 \exp\left(
 \pm\frac{\pi}{4C_{\rm T}} \frac{H}{M_{\rm c}} \Omega
 \right) \ ,
\end{align}
where $\rm U$ is the confluent hypergeometric function.
This solution has the asymptotic form in super-horizon scale:
\begin{align}
 u_{\bf k}(\tau)
 = \sqrt{\frac{-\tau}{2(-C_{\rm T}k\tau)^3}}
 {\rm e}^{i(-\pi/4+\pi\nu_{\rm T}/2)}
 \frac{\Gamma(\nu_{\rm T})}{\Gamma(3/2)}
 \left(\frac{-C_{\rm T}k\tau}{2}\right)^{3/2-\nu_{\rm T}}
 \exp\left(
 \pm\frac{\pi}{4C_{\rm T}} \frac{H}{M_{\rm c}} \Omega
 \right)
\end{align}
We define the power spectra for each polarization mode
${\cal P}^\pm_{\rm T}$ and the total power spectrum
${\cal P}_{\rm T}$: 
\begin{align}
 \langle0| \hat{h}^\dagger_{ij} \hat{h}^{ij} |0\rangle
 = \sum_\pm \int {\rm d}(\log k) \frac{1}{M_{\rm Pl}^2}
 \frac{2}{\pi^2} k^3
 \frac{|u_{\bf k}|^2}{A_{\rm T}^2}
 = \sum_\pm \int {\rm d}(\log k) {\cal P}^\pm_{\rm T}
 = \int {\rm d}(\log k) {\cal P}_{\rm T} \ .
\end{align}
In super-horizon, 
these ${\cal P^\pm_{\rm T}}$ and ${\cal P}_{\rm T}$
become
\begin{align}
 {\cal P}^\pm_{\rm T} &= \frac{1}{M_{\rm Pl}^2}
 \frac{1}{C_{\rm T}^3} \frac{1}{\pi^2} 
 \frac{1}{A_{\rm T}^2\tau^2}
 \left(\frac{\Gamma(\nu_{\rm T})}{\Gamma(3/2)}\right)^2
 \left(\frac{-C_{\rm T}k\tau}{2}\right)^{3-2\nu_{\rm T}}
 \exp\left(
 \pm\frac{\pi}{2C_{\rm T}} \frac{H}{M_{\rm c}} \Omega
 \right)
 \nonumber \\
 &= \frac{1-3\alpha+\beta}{\pi^2}
 \left(
 \frac{H}{M_{\rm Pl}} 
 \right)^2
 \left(\frac{-k\tau}{2}\right)^{3-2\nu_{\rm T}}
 \left(1\pm\frac{\pi}{2}\frac{H}{M_{\rm c}}\Omega\right)
 \label{ptpm}
 \\
 {\cal P}_{\rm T} &= \frac{2}{\pi^2}(1-3\alpha+\beta)
 \left(
 \frac{H}{M_{\rm Pl}} 
 \right)^2
 \left(\frac{-k\tau}{2}\right)^{3-2\nu_{\rm T}} \ .
\end{align}
The tensor spectral index $n_{\rm T}$ is 
\begin{align}
 n_{\rm T} = 3 - 2\nu_{\rm T} = - 2 \alpha \ .
 \label{ntex}
\end{align}
This expression becomes the same as an 
ordinary one in the $\xi\rightarrow 0$ limit.
Remember that in our model, $\alpha$ can takes
a negative value in which super inflation, $\dot{H}>0$, 
is realized. 
Hence, this expression shows that in super inflation, 
a blue spectrum for tensor modes, $n_{\rm T}>0$, 
is realized. 
The detection of this blue signal
might be evidence for this inflationary model,
because $n_{\rm T}<0$ is always satisfied in a conventional
model.

We have derived both scalar and tensor spectra.
Thus,  
we can get the observationally important quantity, namely
the tensor-to-scalar ratio $r$:
\begin{align}
 r \equiv \frac{{\cal P}_{\rm T}}{{\cal P}_{\rm S}}
 = 16 \beta \ .
 \label{rex}
\end{align}
Note that $\beta\rightarrow\epsilon$ in the $\xi\rightarrow 0$
limit.
There is the consistency relation $r=-8n_{\rm T}$
in ordinary single field slow-roll inflation, or
$r\le -8n_{\rm T}$ in multi-filed one.
However, these consistency relation doesn't hold in our
case, because $\alpha$ and $\beta$ are in principle
different quantities. The violation of
this relation might be an observationally 
interesting feature of our model.
Note that a blue spectrum of tensor modes can also
be interpreted
as a strong violation of this relation.

The another important feature of our model is a violation of
parity symmetry.
Eq.(\ref{ptpm}) shows that the parity violating 
Chern-Simons correction
results in difference between power spectra of each
helicity mode. 
We quantify this difference
using the ratio $\Pi$ which is defined by
\begin{align}
 \Pi \equiv \frac{{\cal P}^+_{\rm T} - {\cal P}^-_{\rm T}}
 {{\cal P}^+_{\rm T} + {\cal P}^-_{\rm T}}
 = \frac{\pi}{2} \frac{H}{M_{\rm c}} \Omega \ .
 \label{def_pi}
\end{align}
This quantity $\Pi$
takes naturally a small value, 
because $\Omega < 1$ and $H<M_{\rm c}$. 
However, 
we can still expect $\Pi$ is of order a few percent, which is 
considered to be observable in 
future experiments\cite{Saito:2007kt,Seto:2006hf,taruya}.
Note that  
we have got no observational constraint on 
$\omega$ and $\Omega$, because these only appear in tensor
modes, which we haven't detected. 

In a conventional model, the transformation of potential
$V\rightarrow \rho V$, where $\rho$ is constant,
has no effect on spectral indices, tensor-to-scalar
ratio and e-folding number.
Hence, an overall factor of potential is
 fixed by the scalar power spectrum ${\cal P}_{\rm S}$.
In our model, from Eqs.(\ref{slow-roll_param}) and
(\ref{e-fold_0}), it is clear that 
these observables depend on the transformation
$V\rightarrow\rho V$.
In this case, the transformations which keep 
$\alpha, \beta, \gamma$ and $N$ constants are 
$V\rightarrow\rho V$ and $\xi\rightarrow \xi/\rho$.

\subsection{Reconstruction of potential and coupling functions}
Slow-roll inflation in the Gauss-Bonnet modified gravity
violates the consistency relation $r= -8n_{\rm T}$.
However, in the case of standard multi field inflation,
this relation also modified as $r\le -8n_{\rm T}$.
Therefore, the case of over production of gravitational waves,
$r>-8n_{\rm T}$, is important. 
From Eqs.(\ref{ntex}) and (\ref{rex}), we can
see over-production is
realized if $\beta > \alpha$ is satisfied.
Taking a look at the definition of slow-roll parameters and
$\zeta$, Eqs.(\ref{slow-roll_param}) and (\ref{def_zeta}), 
we can see in the situation that $\xi_{,\phi}>0$ is realized,
gravitational waves will always be overly produced.
Thus, we can expect over production in many models.

If the violation of this relation is observationally
confirmed, we can reconstruct the value of the derivative of 
the Gauss-Bonnet coupling function $\xi_{,\phi}$.
From Eqs.(\ref{slow-roll_param})
\begin{align}
 \frac{\beta}{\alpha} = \frac{\zeta}{V_{,\phi}}
 = 1 + \frac{V^2 \xi_{,\phi}}{6 M_{\rm Pl}^4 V_{,\phi}}
 = 1 + \frac{1}{12 M_{\rm Pl}^2} 
 \frac{\zeta \xi_{,\phi}}{\alpha}
 = 1 + \frac{\sqrt{2}}{12 M_{\rm Pl}^3}
 \frac{\sqrt{\beta}}{\alpha} V \xi_{,\phi} \ .
\end{align}
The potential $V$ and the slow-roll parameter $\beta$ are
related by the scalar power spectrum ${\cal P}_{\rm S}$.
From Eq.(\ref{psex})
\begin{align}
 {\cal P}_{\rm S} = \frac{1}{24\pi^2 \beta} 
 \frac{V}{M_{\rm Pl}^4} \ .
 \label{psex2}
\end{align}
Hence, $\xi_{,\phi}$ is written as
\begin{align}
 \xi_{,\phi} = \frac{12 M_{\rm Pl}^3}{\sqrt{2\beta}}
 \frac{1}{V}
 (\beta-\alpha)
 = \frac{1}{2\sqrt{2}\pi^2 M_{\rm Pl}}
 \frac{1}{{\cal P}_{\rm S}}
 \frac{\beta-\alpha}{\beta^{3/2}}
 = \frac{\sqrt{2}}{\pi^2 M_{\rm Pl}}
 \frac{1}{\sqrt{r}{\cal P}_{\rm S}}
 \frac{r+8 n_{\rm T}}{r} \ .
\end{align}
Let us parameterize the violation ratio
of consistency relation by using $\delta$:
\begin{align}
 \delta \equiv \frac{r+8n_{\rm T}}{r} \ .
\end{align}
The expression for $\xi_{,\phi}$ becomes
\begin{align}
 M_{\rm Pl} \xi_{,\phi} =
 \frac{\sqrt{2}}{\pi^2} 
 \frac{\delta}{\sqrt{r} {\cal P}_{\rm S}}
 = 1.86 \times 10^6 
 \left(\frac{{\cal P}_{\rm S}}{2.441\times 10^{-9}}\right)^{-1}
 \left(\frac{r}{0.1}\right)^{-1/2}
 \left(\frac{\delta}{0.01}\right) \ .
 \label{xiphi}
\end{align}
This equation clearly shows that
we can get the value of $\xi_{,\phi}$,
or constrain $\xi_{,\phi}$ at least,
with observing 
the tensor-to-scalar ratio $r$ and 
the violation ratio of consistency relation $\delta$.
In other words, we can predict this violation ratio of 
consistency relation by using this equation for each
Gauss-Bonnet inflationary model.
The expression for the
 value of potential, at the CMB scale, is
the same as usual. From Eq.(\ref{psex2})
\begin{align}
 \frac{V}{M_{\rm Pl}^4}
 = \frac{3\pi^2}{2}r{\cal P}_{\rm S}
 = 3.61\times 10^{-9}
 \left(\frac{{\cal P}_{\rm S}}{2.441\times 10^{-9}}\right)
 \left(\frac{r}{0.1}\right)
 \ .
 \label{V_o}
\end{align}
However, the derivative of potential $V_{,\phi}$ suffers
modification. From the definition of slow-roll parameters,
Eq.(\ref{slow-roll_param})
\begin{align}
 \frac{V_{,\phi}}{M_{\rm Pl}^3}
 &= \frac{2}{M_{\rm Pl}^8} \frac{V}{\zeta}V\alpha
 = \frac{\sqrt{2}\alpha V}{M_{\rm Pl}^4 \sqrt{\beta}}
 = 24\sqrt{2}\pi^2 \alpha \sqrt{\beta} {\cal P}_{\rm S}
 = -3\sqrt{2} \pi^2 n_{\rm T} \sqrt{r} {\cal P}_{\rm S}
 = \frac{3\sqrt{2}\pi^2}{8} r^{3/2} {\cal P}_{\rm S}
 (1-\delta)
 \nonumber
 \\
 &= 4.04 \times 10^{-10} \left(
 \frac{{\cal P}_{\rm S}}{2.441\times 10^{-9}}
 \right)
 \left(\frac{r}{0.1}\right)^{3/2}
 (1-\delta) \ .
 \label{dV_o}
\end{align}

Using Eqs.(\ref{xiphi}),(\ref{V_o}) and (\ref{dV_o}),
we can fix the value of $V,\ V_{,\phi}$ and $\xi_{,\phi}$
using observables.
The expression for the scalar spectral index $n_{\rm S}$ contains
$V_{,\phi\phi}$ and $\xi_{,\phi\phi}$. 
However, we cannot separate these quantities in our model.
Thus, from these observables, we cannot fix the value
of $V_{,\phi\phi}$ and $\xi_{,\phi\phi}$.
To this point, we have studied a general case, 
placing special emphasis on a
violation of consistency relation. 
However, there are
other models which also can violate this 
relation. Particularly, the model which
overly produce tensor modes exists\cite{mukhanov06}.
Therefore in the 
next section, we will investigate an unique
phenomenon in our model, namely a blue spectrum of 
tensor mode.

Next, let us study the Chern-Simons coupling function.
Considering future observations,
deriving the expression for $\omega_{,\phi}$ by using
observable is meaningful.
From the definition of $\Omega$,
Eq.(\ref{def_Omega}), $\omega_{,\phi}$ is
\begin{align}
 M_{\rm Pl}\omega_{,\phi} &= 
 M_{\rm Pl}\frac{\dot{\omega}}{\dot{\phi}}
 = \frac{2M_{\rm Pl}^3\Omega}{M_{\rm c}}
 \frac{-1}{\sqrt{2\beta}M_{\rm Pl}H}
 = - \frac{4M_{\rm Pl}^2}{\pi} \frac{\Pi}{\sqrt{2\beta}H^2}
 = - \frac{12M_{\rm Pl}^4}{\pi}
 \frac{\Pi}{\sqrt{2\beta}V}
 = - \frac{\sqrt{2}}{4\pi^3} 
 \frac{\Pi}{\beta^{3/2}{\cal P}_{\rm S}}
 = - \frac{16\sqrt{2}}{\pi^3} 
 \frac{\Pi}{r^{3/2} {\cal P}_{\rm S}}
 \nonumber \\
 &= - 9.45\times 10^{7}
 \left(\frac{{\cal P}_{\rm S}}{2.441\times 10^{-9}}\right)^{-1}
 \left(\frac{r}{0.1}\right)^{-3/2}
 \left(\frac{\Pi}{0.01}\right) \ .
\end{align}
Thus, we can see that
for confirming existence of the Chern-Simons term, 
it is important to
detect circular polarization in tensor modes.
We can also use this expression to predict this polarization ratio.
Because $|\Pi|\ll1$ is required by the definition,
Eq.(\ref{def_pi}),
we can get the constraint on the value of 
$\omega_{,\phi}$:
\begin{align}
 |M_{\rm Pl} \omega_{,\phi}| 
 \ll 9.45 \times 10^9
 \left(\frac{{\cal P}_{\rm S}}{2.441\times 10^{-9}}\right)^{-1}
 \left(\frac{r}{0.1}\right)^{-3/2} \ .
\end{align}

Of course, if we allow an appearance of ghost in tensor
modes, a situation might be changed drastically. However,
it exceeds the scope of this paper.

\subsection{Lyth bound}
Here, we comment on the Lyth bound\cite{lyth97}.
The statement of this bound is following:
If primordial gravitational
waves are detected, the variation of inflaton field
must exceed a scale on the order of Planck mass.
In our model, this bound also holds.
From Eq.(\ref{e-fold_0})
\begin{align}
 N = \frac{1}{M_{\rm Pl}}
 \int^\phi_0 \frac{{\rm d}\phi}{\sqrt{2\beta}}
 = \frac{2\sqrt{2}}{M_{\rm Pl}}
 \int^\phi_0 \frac{{\rm d}\phi}{\sqrt{r}} \ ,
\end{align}
and
\begin{align}
 \delta\phi = \frac{M_{\rm Pl}}{2}
 \sqrt{\frac{r}{2}} \delta N \ .
\end{align}
This can be easily understood: Large $r$ corresponds to small 
$\dot{\phi}$, and this leads to small $\delta \phi$.
Note that effect of gravitational waves on the CMB multi pole
is relevant in approximately $l\lesssim100$, which corresponds
to $\delta N\simeq 4$.
This means that if gravitational waves with tensor-to-scalar
ratio $r$ are observed, the field value of inflaton has
changed its value
approximately $\sqrt{2r}$ during $\delta N\simeq 4$.
Hence, from this equation, we can set the lower bound
on variation of inflaton field $\Delta\phi$ during
whole inflation:
\begin{align}
 \frac{\Delta \phi}{M_{\rm Pl}} \gtrsim \sqrt{2r}
 = 0.447
 \left(\frac{r}{0.1}\right)^{1/2}
 \ .
\end{align}
We can see that the Lyth bound holds in our case,
while an equation of motion is modified due to the
Gauss-Bonnet correction.
Therefore, extremely small field inflation, namely
$\Delta\phi\ll M_{\rm Pl}$, cannot produce
an observable amount of tensor modes.

\section{Interesting features of our model}
In our model, a spectrum of primordial gravitational
waves can be blue or scale invariant, and these cases strongly
violate the consistency relation $r=-8n_{\rm T}$.
Hence, these (especially, a blue spectrum) 
are important for confirming our model with
future observations.
In this section, we use concrete examples to
check these interesting situations are consistent
with the current WMAP result.
We show characteristics of potential and the Gauss-Bonnet 
coupling function
which realize a blue or scale invariant spectrum.
We also consider
the case in which potential is cancelled out by
the Gauss-Bonnet effective potential.

We take e-folding number as $N=60$ for
calculations in examples.

\subsection{Blue spectrum}
\begin{figure}[htbp]
 \centering
 \includegraphics[scale=0.7]{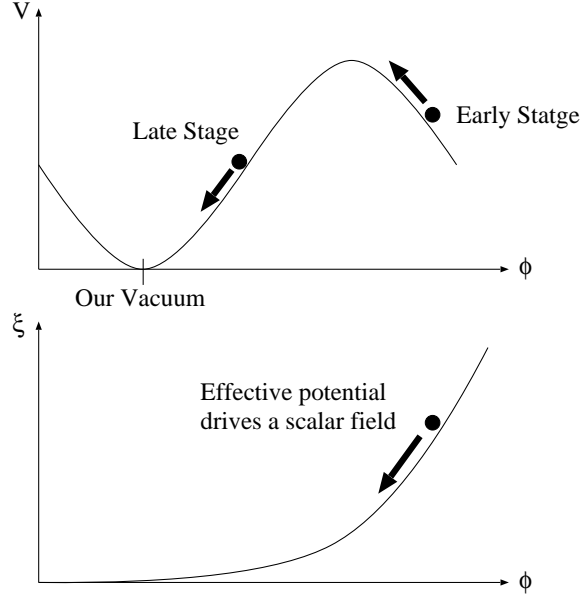}
 \caption{The configuration of the potential $V$ and 
 the coupling function $\xi$, which achieve a 
 blue spectrum, is shown.
 We need a climbing-up situation to realize a blue spectrum.
 Therefore the following situation is required:
 At early stage, the effective potential $\xi$ makes 
 $\phi$ climb up potential and a blue spectrum is realized.
 At late stage, $\phi$ rolls down potential as usual.}
 \label{blue}
\end{figure}

The most interesting and unique 
phenomenon in our model might be
a blue spectrum of tensor modes.  
Note that it is impossible to achieve a blue spectrum in a
conventional model.
For achieving a blue spectrum, $\alpha<0$ is required.
Hence $V_{,\phi}<0$ is necessary because $\dot{\phi}<0$.
In a case of 
blue spectrum, a scalar field is climbing up 
potential, not rolling down. 
Of course, this blue situation must be over at some time,
or inflation lasts forever, because $\dot{H}>0$ is satisfied 
in this blue case. 
A climbing up situation can be understood 
using Eq.(\ref{eom_slow}), 
which shows that the coupling function $\xi$ works as 
effective potential.
Hence, if the effective potential $\xi$ is steep enough, 
$\phi$ can climbs up potential.
We show this in FIG.\ref{blue}.

In a blue spectrum case, $\xi_{,\phi}$ can be
calculated by employing Eq.(\ref{xiphi}) with $\delta > 1$.
Hence, there is a necessary condition for a blue spectrum:
\begin{align}
 M_{\rm Pl}\xi_{,\phi}>\frac{\sqrt{2}}{\pi^2}
 \frac{1}{\sqrt{r}{\cal P}_{\rm S}}
 = 1.86\times 10^8
 \left(\frac{{\cal P}_{\rm S}}{2.441\times 10^{-9}}\right)^{-1}
 \left(\frac{r}{0.1}\right)^{-1/2} \ ,
 \label{blueconst0}
\end{align}
and combining with Eq.(\ref{V_o})
\begin{align}
 \frac{V\xi_{,\phi}}{M_{\rm Pl}^3}
 > \frac{3}{\sqrt{2}}\sqrt{r}
 = 0.671 \left(\frac{r}{0.1}\right)^{1/2} \ .
 \label{blueconst}
\end{align}
Remember $\delta=1+8n_{\rm T}/r$ and $|n_{\rm T}|\ll1$. 
To observe tensor modes,
$r$ cannot take an extremely small value.
Hence, 
$\delta$ cannot exceed a value of order unity
and $\delta\sim1$ must hold.
In a consistent blue spectrum case, hence, 
we need the derivative of the
Gauss-Bonnet coupling function $\xi_{,\phi}$ on the 
order of $10^8/M_{\rm Pl}$ and $M_{\rm Pl}^3/V$. 
It is important that both too small and too large
a value of $\xi_{,\phi}$ cannot result in a desirable blue
spectrum. In other words, if a blue spectrum is detected,
$\xi_{,\phi}$ at the CMB scale must be on the order of 
$10^8/M_{\rm Pl}$.

In the case that a blue spectrum of tensor modes is confirmed, 
we can reconstruct the value of $\xi_{,\phi}$, which
must be on the order of $10^8/M_{\rm Pl}$.
However, 
it is unclear whether or not
an observationally consistent
 blue spectrum case can actually be realized;
what kind of potential and the Gauss-Bonnet coupling function
is likely to realize a blue spectrum.
To show these, 
we use two types of examples of blue spectra. 
First one is a simple, but analytically solvable
model and second one is a somewhat realistic model, which
is calculated numerically.

We take the following potential and coupling function:
\begin{align}
 V(\phi) = V_0
 \left| \sin
 \left( \frac{\pi}{2}\frac{\phi}{\phi_0} \right)
 \right| \ , \quad
 \xi_{,\phi}(\phi) = {\rm const.}
\end{align}
In this case, the 
e-folding number $N$ can be written analytically:
\begin{align}
 N = \frac{4 \phi_0^2}{\pi^2 M_{\rm Pl}^2}
 \frac{1}{\sqrt{1+4\theta^2}}
 \ln\left(\frac{x_1-x}{x-x_2}\frac{1-x_2}{x_1-1}\right)
 \ ,
\end{align}
where
\begin{align}
 x \equiv \cos\left(\frac{\pi}{2}\frac{\phi}{\phi_0}\right)
 \ , \quad
 \theta \equiv \frac{V_0 \xi_{,\phi} \phi_0}{3\pi M_{\rm Pl}^4}
 \ , \quad
 x_1 \equiv \frac{1+\sqrt{1+4\theta^2}}{2\theta}
 \ , \quad
 x_2 \equiv \frac{1-\sqrt{1+4\theta^2}}{2\theta} 
 = -\frac{1}{x_1}
 \ .
\end{align}
Note that $x$ must be in the range $-1<x_2<x<1<x_1$ and
$\theta$ represents effect of the Gauss-Bonnet correction.
We use $x$ as a variable instead of $\phi$ and
$x<0$ corresponds to 
a climbing-up situation and
a blue spectrum. To show a realization of
blue spectrum, we express
$x$ in $\theta$ and $N$:
\begin{align}
 x = 
 \frac{1+x_1 - (1+x_2){\rm e}^y}{1-x_2 - (1-x_1){\rm e}^y}
 = \frac{
 \sqrt{1+4\theta^2}+1+2\theta + (\sqrt{1+4\theta^2}-1-2\theta)
 {\rm e}^y
 }{
 \sqrt{1+4\theta^2}-1+2\theta + (\sqrt{1+4\theta^2}+1-2\theta)
 {\rm e}^y
 } \ ,
 \label{x0}
\end{align}
where
\begin{align}
 y \equiv 
 \frac{\pi^2M_{\rm Pl}^2}{4\phi_0^2}
 \sqrt{1+4\theta^2} N
 = y_0 \sqrt{1+4\theta^2}
 \ .
 \label{defy}
\end{align}
Here, we have two parameters in this model, namely
$\theta$ and $y_0$.
Let us consider varying $\theta$ to realize a blue spectrum
in fixed $y_0$, which corresponds to fixed $\phi_0$.
In the situation that $\theta\gg1$,  Eq.(\ref{x0}) becomes
\begin{align}
 x = \frac{4\theta - {\rm e}^{2y_0\theta}}
 {4\theta + {\rm e}^{2y_0\theta}} \ .
\end{align}
Hence, with sufficiently large $\theta$,
$x<0$ and blue spectrum can be realized for any $\phi_0$
value.
The condition for a blue spectrum is
\begin{align}
 4 (1-q) \theta^2 - 4q \theta + (1-q) < 0 \ ,
 \label{thetal0}
\end{align}
where
\begin{align}
 q \equiv \tanh^2\left(\frac{y}{2}\right) \ .
 \label{defq}
\end{align}
This inequality is satisfied, only if $1/2<q<1$.
We define $y_{\rm b}$ so that $x=0$ 
is realized at $y=y_{\rm b}$.
Note that $y < y_{\rm b}$ and $y > y_{\rm b}$
correspond to red and blue spectra, respectively.
Because $1/2<q$ is necessary for a blue spectrum,
the condition for $y_{\rm b}$ is
\begin{align}
 y_{\rm b} >  y_{\rm c}
 \equiv 2 \tanh^{-1} \frac{1}{\sqrt{2}} = 1.76 \ .
 \label{cony}
\end{align}
Here, we define $q_{\rm b}$ and 
$\theta_{\rm b}$ in the same manner as
$y_{\rm b}$.
From Eqs.(\ref{defy}), (\ref{thetal0}) and (\ref{defq}), 
we can relate these variables:
\begin{align}
 q_{\rm b} = 
 \frac{4\theta_{\rm b}^2 + 1}{(2\theta_{\rm b}+1)^2} = 
 \tanh^2\left(\frac{y_{\rm b}}{2}\right)
 =
 \tanh^2\left(\frac{y_0\sqrt{1+4\theta_{\rm b}}}{2}\right)
 \ .
\end{align}
Note that for $y_0=y_{\rm c}/\sqrt{2}$ model,
$\theta_{\rm b}=1/2$ (and $\ y_{\rm b}=y_{\rm c}$) 
is the solution for above equations.
The scalar spectral index $n_{\rm S}$, 
the tensor spectral index $n_{\rm T}$ and
the tensor-to-scalar ratio $r$ become
\begin{align}
 n_{\rm S}-1 &=
 \frac{y_0}{N} \frac{1}{1-x^2}
 \left[
 -2-x^2 + \theta x(1-x^2)
 \right] \\
 n_{\rm T} &=
 \frac{y_0}{N} \frac{-x}{1-x^2}
 \left[
 x + \theta (1-x^2)
 \right]
 =
 \frac{y_0}{N} \frac{\theta x}{1-x^2}
 (x-x_1)(x-x_2) \\
 r &=
 \frac{8 y_0}{N} \frac{1}{1-x^2}
 \left[x + \theta (1-x^2)\right]^2
 =
 \frac{8 y_0}{N} \frac{\theta^2}{1-x^2}
 (x-x_1)^2 (x-x_2)^2
 \ .
\end{align}

Let us consider the large $\theta$ limit, in which
$y \gg y_{\rm b}$ is satisfied.
In this limit, the following equation is realized:
\begin{align}
 x-x_2 = 
 \frac{1+x_1-x_2+x_2^2}{1-x_2-(1-x_1){\rm e}^y}
 \rightarrow 0 \ .
\end{align}
Hence, $x$ approaches to $x_2$.
We define new variable $z$:
\begin{align}
 z \equiv x + \theta(1-x^2) = - \theta(x-x_1)(x-x_2) \ .
\end{align}
Because $x\rightarrow x_2$ is realized, $z\rightarrow 0$ holds. 
Note that $x-x_2$ approaches to zero
exponentially as a function of  $\theta$, and this leads to
$\theta(x-x_2)\rightarrow0$. 
Observables become
\begin{align}
 n_{\rm S} -1 = -\frac{4y_0}{N}\theta \ , \quad
 n_{\rm T} = \frac{y_0}{N}\theta z \ , \quad
 r = \frac{8y_0}{N}\theta z^2 \ .
\end{align}
Note that $r$ becomes small in large $\theta$, 
because $\theta z\rightarrow 0$.
This shows that 
in any model, this large $\theta$ limit results in
small $r$ and extremely red $n_{\rm S}$.
The reason for small $r$ is that
$x=x_2$ corresponds to the fixed point, at which potential is
cancelled out by effective potential. Hence, 
$\dot{\phi}=0$ and $r=0$ are realized.
Since this is not a desirable case,
we consider the small $\theta$ case, in which
$\theta\sim\theta_{\rm b}$ holds.
In this case, $x\sim 0$ is realized, and observables become
\begin{align}
 n_{\rm S} - 1 =
 -\frac{2y_0}{N} \ ,\quad
 n_{\rm T} = 
 \frac{y_0}{N}(-x\theta_{\rm b})\ , \quad
 r = \frac{8y_0}{N}\theta_{\rm b}^2 \ .
 \label{ob0}
\end{align}
Note that $\theta_{\rm b}$ and $y_0$ are not independent
variables, because
 $\theta_{\rm b}$ is fixed by choosing $y_0$.
Remember that $\theta_{\rm b}=1/2$ is the solution for
$y_0=y_{\rm c}/\sqrt{2}$ model.
Let us substitute these:
\begin{align}
 n_{\rm S} - 1 = -\frac{\sqrt{2}y_{\rm c}}{N}
 = -0.0415 \ ,\quad
 n_{\rm T} = \frac{y_{\rm c}}{2\sqrt{2}N}(-x) 
 = -0.0104 x\ ,\quad
 r = \frac{\sqrt{2}y_{\rm c}}{N}
 = 0.0415 \ .
\end{align}
Here, we get good result: $n_{\rm S}$ and $r$ are
consistent with current observations and a blue 
spectrum of tensor modes is obtained.
Therefore, desirable solutions might exist in the
vicinity of $\theta\sim1/2$ and $y_0\sim y_{\rm c}/\sqrt{2}$:
\begin{align}
 \phi_0 \sim \frac{2^{1/4}\pi}{2}
 \sqrt{\frac{N}{y_{\rm c}}}
 M_{\rm Pl}
 = 10.9 M_{\rm Pl}
 \ , \quad
 \frac{V_0\xi_{,\phi}}{M_{\rm Pl}} \sim 
 \frac{3}{2^{1/4}}\sqrt{\frac{y_{\rm c}}{N}}
 = 0.432
 \ .
\end{align}
Let us consider the model, in which $\phi_0$ is large 
enough to satisfy $y_0\ll y_{\rm c}$. 
From Eqs.(\ref{defy}) and (\ref{cony}),
$y_{\rm b}=2y_0 \theta_{\rm b}$ is realized, and $r$ at
$x=0$ becomes
\begin{align}
 r = \frac{4}{N}\theta_{\rm b} y_{\rm b}
 > \frac{4y_{\rm c}}{N}\theta_{\rm b} \ .
\end{align}
Because $\theta_{\rm b}=y_{\rm b}/2y_0> y_{\rm c}/2y_0 \gg1$,
this large field model results in $r$, which is too large.
This is because, in this large field model, inflaton $\phi$ must 
roll large field value between
the hill-top of potential and the vacuum, within $N=60$
e-fold. This requires large $\dot{\phi}$, in other words,
large $r$.
In addition, from Eq.(\ref{ob0}), 
this large field model also leads
to a flat spectrum of scalar modes.
In the small field model, which satisfies 
$y_0\gg y_{\rm c}$, the inequality $y\gg1$ always holds.
Hence, as in the above large $\theta$ limit, 
this results in small $r$ and 
extremely red $n_{\rm S}$. 
Note that larger field models are corresponding to 
larger values of $r$, and
this is consistent with the consideration in the Lyth bound.

\begin{figure}[htbp]
 \centering
 \includegraphics[scale=0.6]{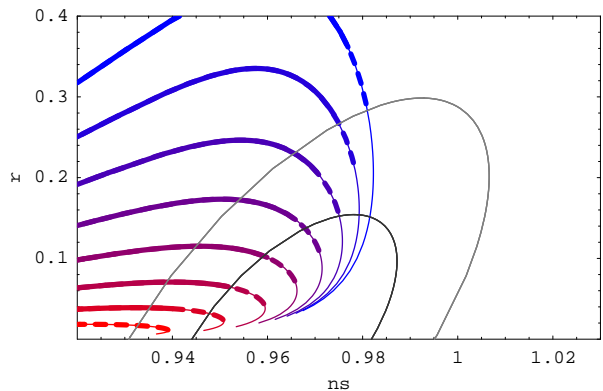}
 \caption{The trajectories, in which $\theta$ is varying,
 of analytic blue models are
 shown on a $n_{\rm S}-r$ plane.
 Two contours roughly corresponds to the 
 68\% and 95\% confidence level of the WMAP 7-year 
 result.
 We draw eight lines, which are corresponding to
 $\phi_0=9$ (red, bottom) to $16M_{\rm Pl}$ (blue, top) 
 with an even interval. 
 Thick dotted and thick solid lines denote the blue spectra of 
 $0.003>n_{\rm T}>0$ and $n_{\rm T}>0.003$, respectively.
 Models with $\phi_0\sim 11M_{\rm Pl}$ are consistent with 
 observations and realizes blue spectra.
 }
 \label{trjsin}
\end{figure}

We show $\theta$-trajectories of solutions on 
a $n_{\rm S}-r$ plane in FIG.\ref{trjsin}.
We plot the 68\% and 95\% C.L.
contours from the WMAP 7-year result\cite{komatsu10}. 
We choose eight $\phi_0$ values
surrounding $\phi_0=11M_{\rm Pl}$. Thick dotted
and thick solid lines
correspond to $0.003>n_{\rm T}>0$ and $n_{\rm T}>0.003$
blue spectra.
As stated above, models with $\phi_0\sim 11M_{\rm Pl}$ 
have observationally consistent blue regions; 
large field models correspond to 
large $r$ and small or no consistent blue spectrum region;
small field models have extremely small $r$; 
in large $\theta$ limit, $r$ becomes zero and $n_{\rm S}$ 
becomes extremely red. 
It is clear that a consistent blue spectrum model can 
actually be constructed.
To get the value of $\xi_{,\phi}$, we
can use Eq.(\ref{xiphi}). Of course, as mentioned above,
$\xi_{,\phi}\sim 10^8/M_{\rm Pl}$ is realized in a blue region.

Next, we consider a realistic model.
To realize a blue spectrum, 
new-inflation-type potential, as depicted in FIG.\ref{blue},
is required.
Therefore, we use forth-order double-well
potential and an exponential coupling function:
\begin{align}
 V(\phi) =  
 \rho 10^{-10} M_{\rm Pl}^4
 \left[
 \frac{(\phi - \phi_0)^2 - \phi_0^2}{\phi_0^2}
 \right]^2
 \ , \quad
 \xi(\phi)
 = s \frac{b}{\rho} \exp(s a \phi/M_{\rm Pl}) \ ,
\end{align}
where $s=\pm1$,
$\rho$ will be fixed by the scalar power spectrum
${\cal P}_{\rm S} = 2.441 \times 10^{-9}$\cite{komatsu10}
and $a, b>0$.
We take $\phi=0$ as our vacuum.
We use this exponential coupling,
since it appears in many theories,
such as a dilatonic case or an one-loop correction from
a superstring motivated model\cite{Antoniadis:1993jc}.

\begin{figure}[htbp]
 \centering
 \begin{minipage}{0.49\hsize}
  \centering
  \includegraphics[scale=0.6]{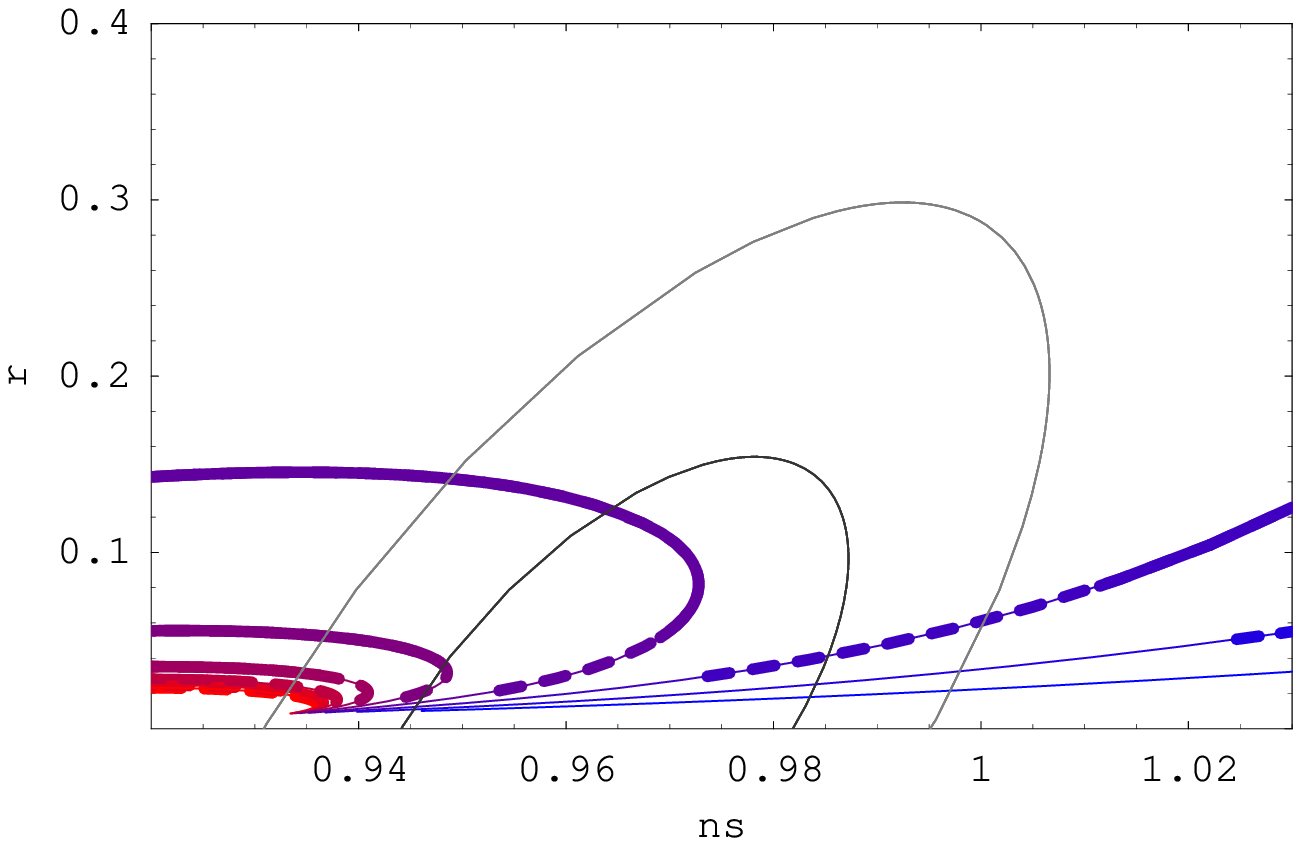}
  $\phi_0=11.1 M_{\rm Pl},\ s=+1$
  \vspace*{0.3cm}
 \end{minipage}
 \begin{minipage}{0.49\hsize}
  \centering
  \includegraphics[scale=0.6]{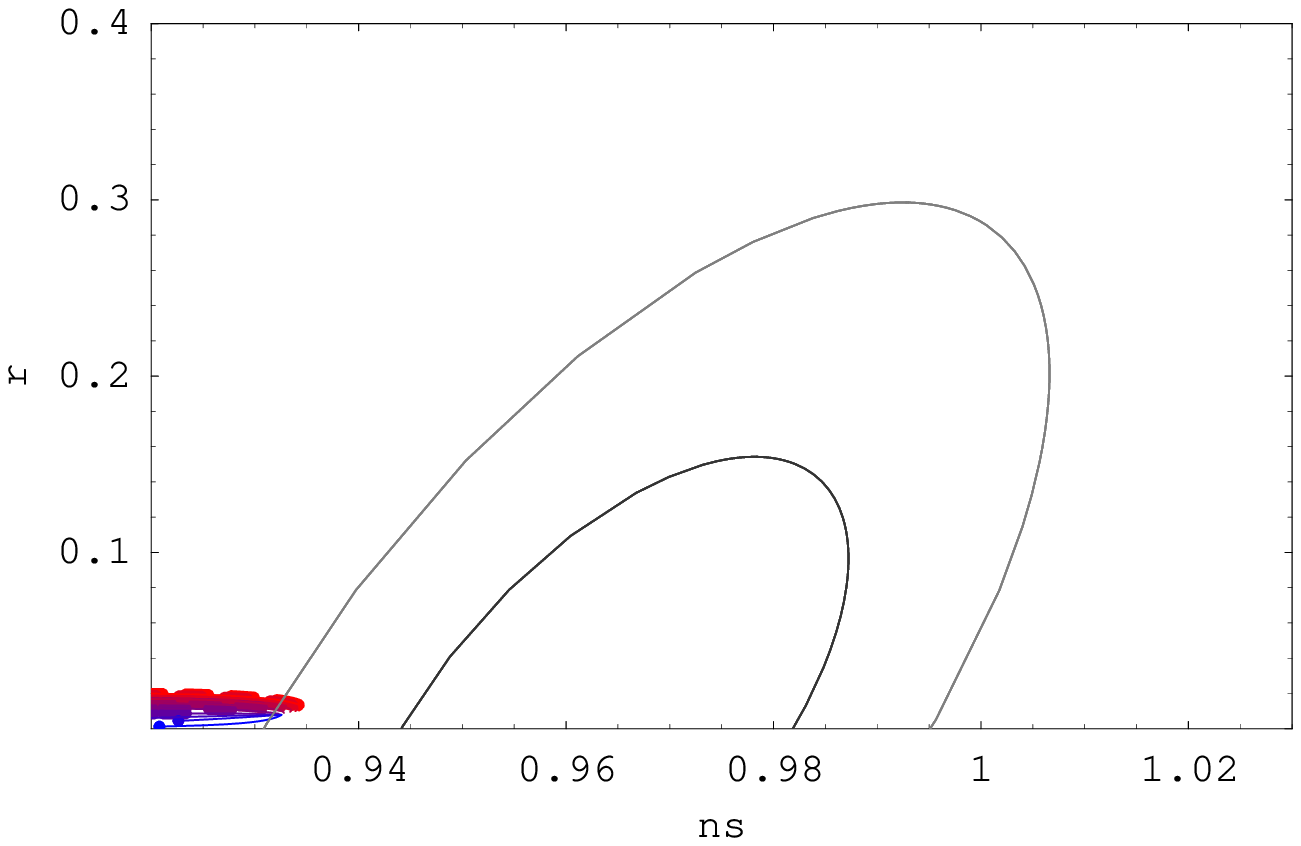}
  $\phi_0=11.1 M_{\rm Pl},\ s=-1$
  \vspace*{0.3cm}
 \end{minipage}
 \begin{minipage}{0.49\hsize}
  \centering
  \includegraphics[scale=0.6]{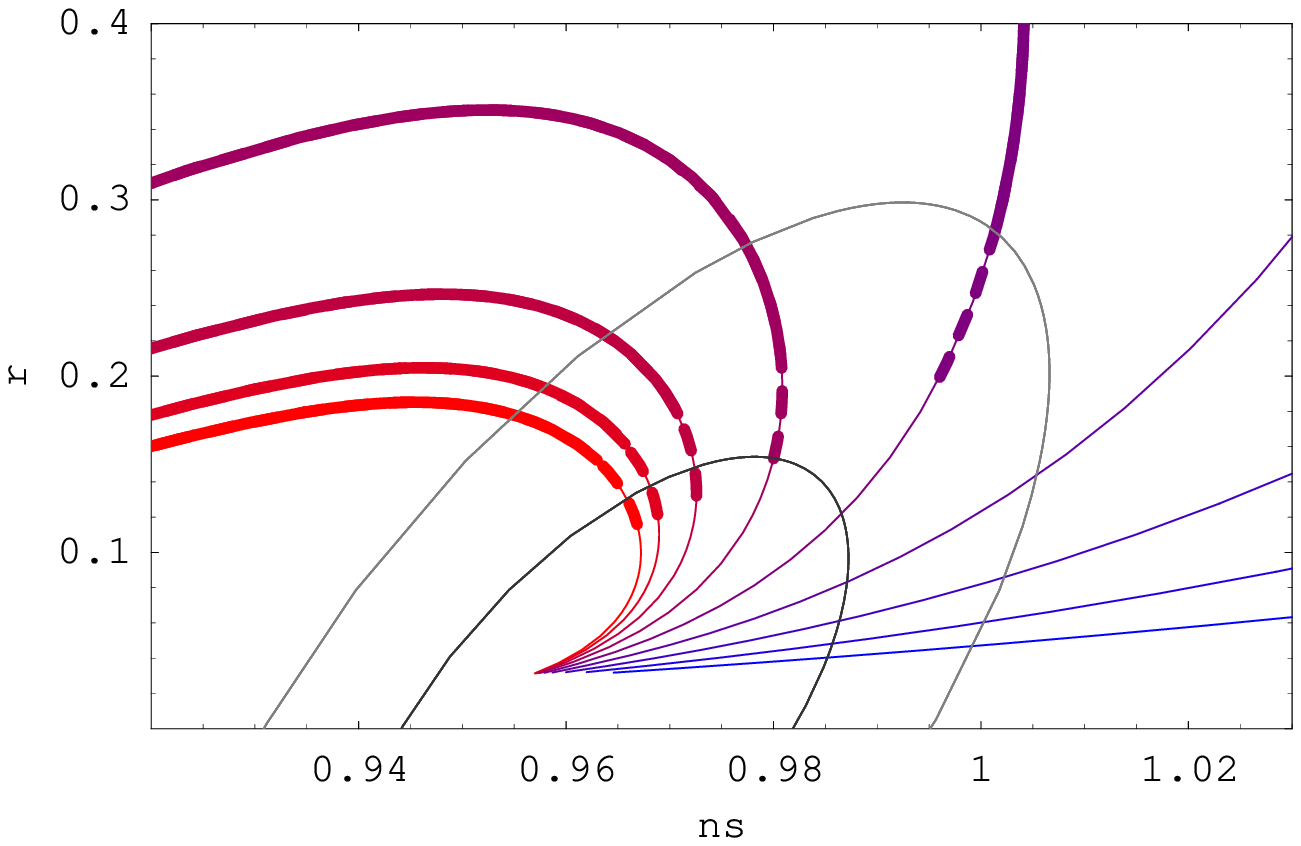}
  $\phi_0=15 M_{\rm Pl},\ s=+1$
  \vspace*{0.3cm}
 \end{minipage}
 \begin{minipage}{0.49\hsize}
  \centering
  \includegraphics[scale=0.6]{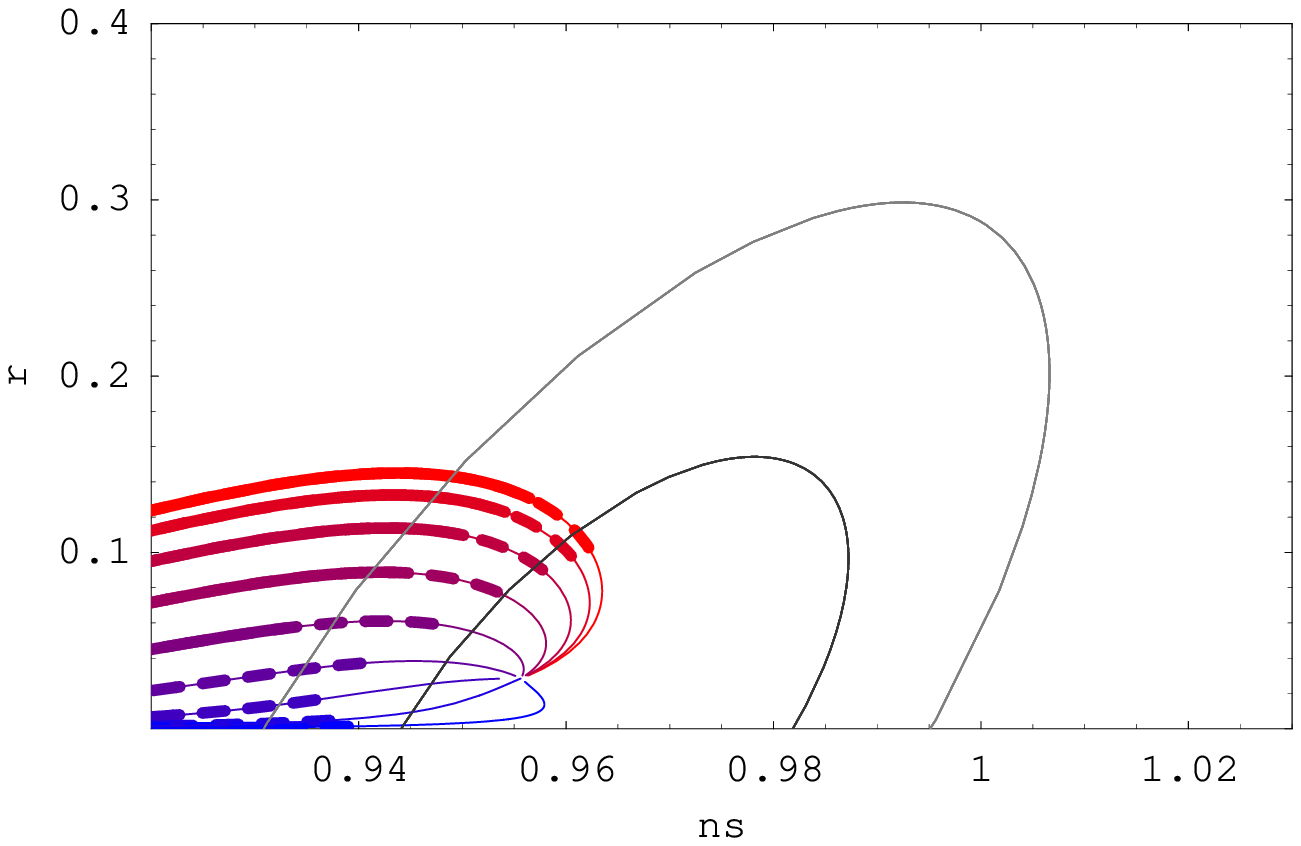}
  $\phi_0=15 M_{\rm Pl},\ s=-1$
  \vspace*{0.3cm}
 \end{minipage}
 \begin{minipage}{0.49\hsize}
  \centering
  \includegraphics[scale=0.6]{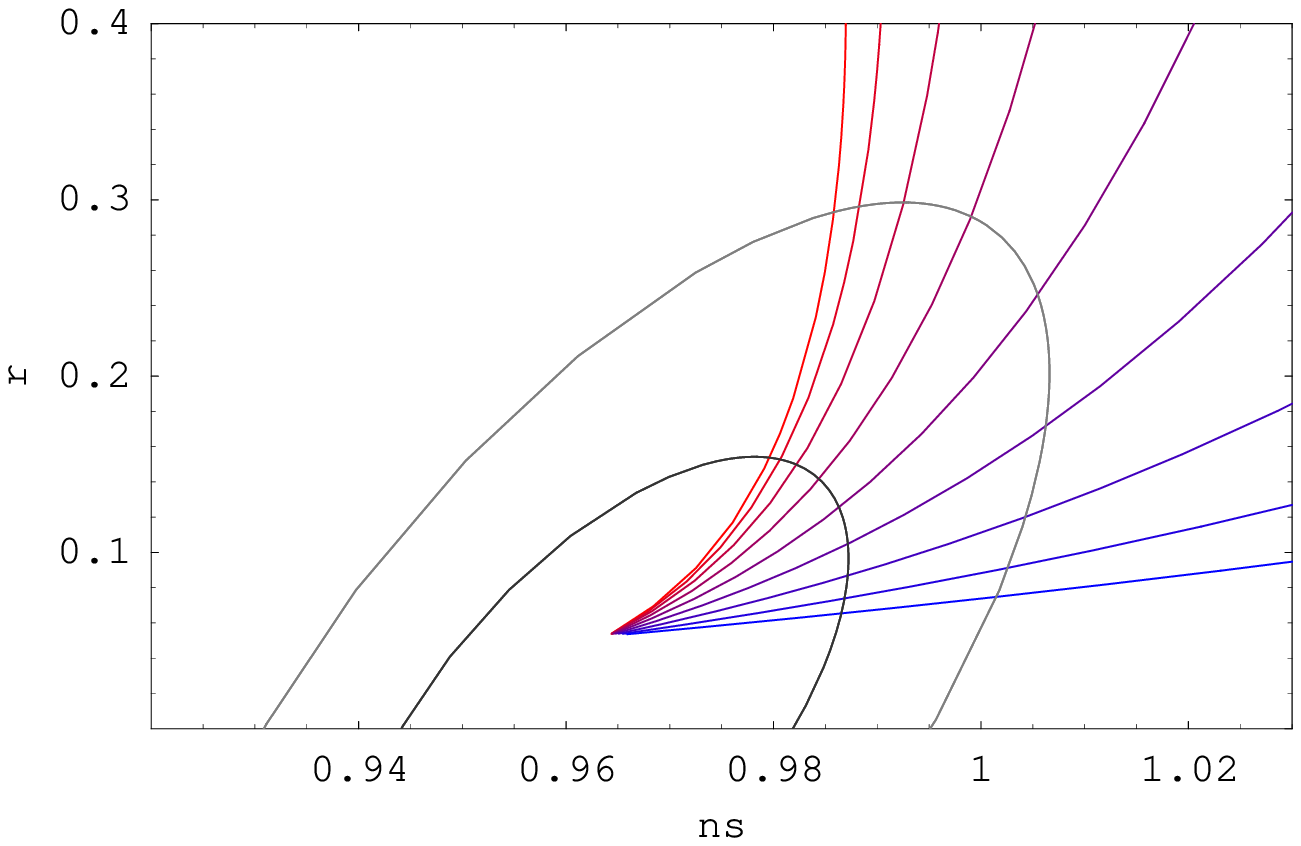}
  $\phi_0=20 M_{\rm Pl},\ s=+1$
 \end{minipage}
 \begin{minipage}{0.49\hsize}
  \centering
  \includegraphics[scale=0.6]{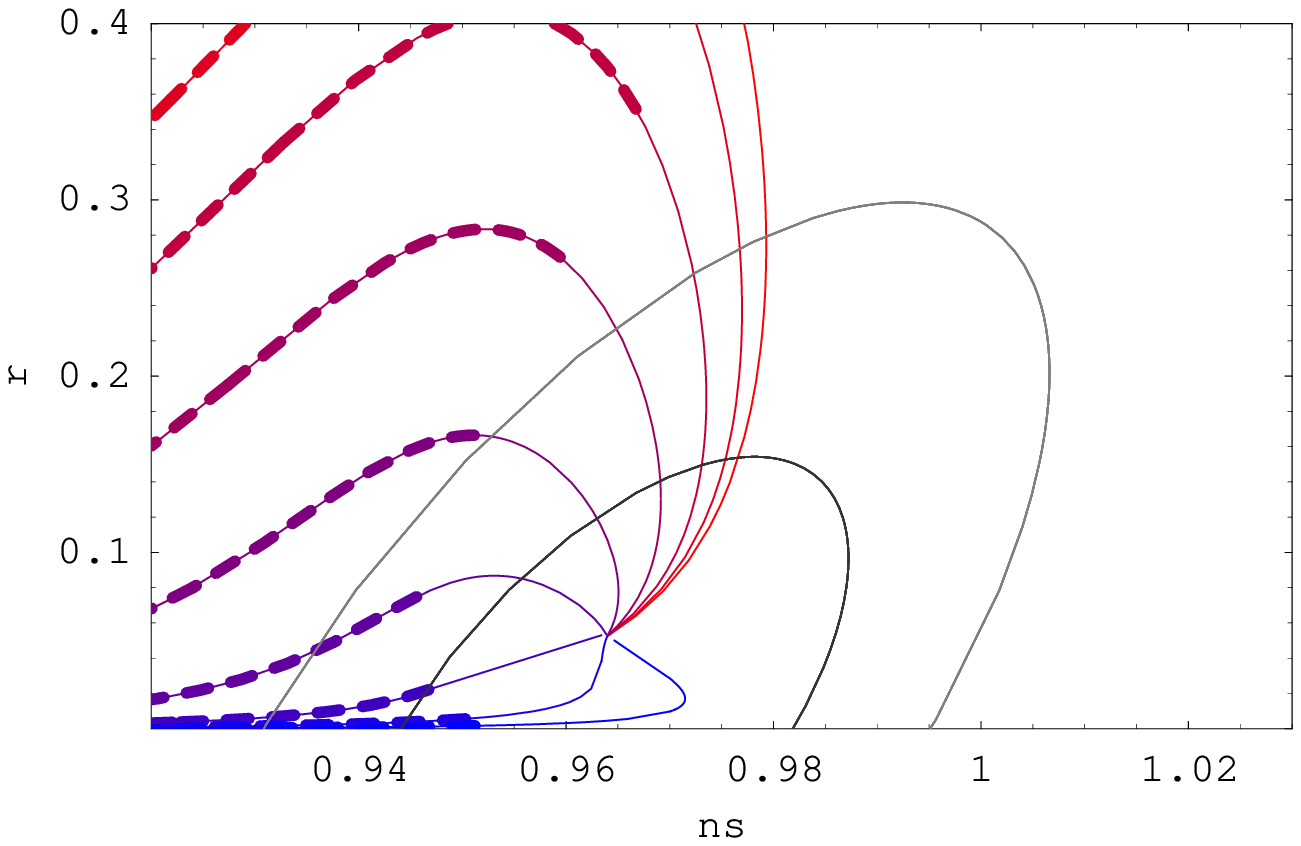}
  $\phi_0=20 M_{\rm Pl},\ s=-1$
 \end{minipage}
 \caption{The trajectories, in which $b$ is varying,
 of some blue models  are
 shown on a $n_{\rm S}-r$ plane.
 In the $b\rightarrow0$ limit, all trajectories converges
 to one point.
 Two contours roughly corresponds to the 
 68\% and 95\% confidence level of the WMAP 7-year 
 result.
 We draw nine lines, which are corresponding to
 $a=0.01$ (red, left in left panels and top in right panels) 
 to $1$ (blue, right in left panels and bottom in right panels)
 with an even logarithmic
 interval. Thick dotted and thick solid
 lines denote $0.003>n_{\rm T}>0$ and $n_{\rm T}>0.003$
 blue spectra, respectively.
 The model with $\phi_0=15M_{\rm Pl}$ is appropriate to achieve
 blue spectrum.
 An exponential coupling
 makes large $\phi_0$ ($\phi_0=20M_{\rm Pl}$) model 
 result in  a blue spectrum, and
 small $\phi_0$ ($\phi_0=11.1M_{\rm Pl}$) model achieve
 an observable amount of $r$.
 }
 \label{trj1a}
\end{figure}

We show $b$-trajectories on a $n_{\rm S}-r$ plane
in FIG.\ref{trj1a}.
We plot the 68\% and 95\% C.L.
contours from the WMAP 7-year result.
We use nine values of $a$, which varies
from $0.01$ to $1$ with an even logarithmic
interval, and three value of $\phi_0$, 
$11.1,\ 15$ and $20M_{\rm Pl}$.
We choose $\phi_0$ value so that $n_{\rm S}$
and $r$ values are inside this 68\% C.L. contour
in the $b\rightarrow0$ limit.
Thick dotted and thick solid lines in FIG.\ref{trj1a}
denote the blue spectra of $0.003>n_{\rm T}>0$ and 
$n_{\rm T}>0.003$, respectively. 
We can see it is possible to realize 
a consistent blue spectrum.
The case 
of $\phi_0=15M_{\rm Pl}$ is appropriate to
achieve a consistent
blue spectrum. Small $\phi_0$ leads to small $r$
and large $\phi_0$ leads to a small consistent blue region, 
and these are essentially 
consistent with the above analytic case.
Due
to this exponential coupling,
even
large field models, 
namely $\phi_0=20M_{\rm Pl}$ models with $s=-1$,
have blue 
spectrum regions and
small field models,
namely $\phi_0=11.1M_{\rm Pl}$ models with $s=+1$, 
have an observable amount of $r$.
This figure shows that an order unity value of
$a$ leads to a non-desirable result, which has
too small $r$ or no blue region.
Therefore $a\ll 1$ is appropriate to obtain a blue spectrum.
Moreover, if $a$ is large in $s=-1$ models, it implies
the Gauss-Bonnet effect becomes stronger toward the end of 
inflation. 
In this case, standard reheating scenarios seem to be broken. 
Therefore, large $a$ in $s=-1$ models is also not
desirable in this sense.
Note that with Eq.(\ref{xiphi}), we can get the value of 
$\xi_{,\phi}$.

From the above analytic model and this realistic one,
we can say that inflationary models with
observationally consistent blue spectra
can be constructed.
Hence, the Gauss-Bonnet coupling function will 
actually be 
fixed or constrained with future observations.
New-inflation-type potential with the symmetry breaking scale of 
$\phi_0 \sim 10 M_{\rm Pl}$ and an almost constant form
of $\xi_{,\phi}$
might lead to a consistent blue spectrum.
Of course, these are not general conditions.
However, we can expect many models are also consistent with
these conditions.
Note that to obtain blue spectrum,
$\xi_{,\phi}\sim 10^8/M_{\rm Pl}$ must be realized in any model
at the CMB scale.

\afterpage{\clearpage}
\subsection{Scale invariant spectrum}
In inflation with a standard gravitational action, 
a scale invariant
spectrum of gravitational waves corresponds to almost purely
de Sitter inflation and this requires the tensor-to-scalar
ratio $r$ extremely small.
However, inflation with the Gauss-Bonnet term behaves
differently.
Let us consider the case, in which the Gauss-Bonnet term
dominate the scalar equation of motion, Eq.(\ref{dphizeta}).
The function $\zeta$ becomes
\begin{align}
 \zeta = \frac{V^2\xi_{,\phi}}{6M_{\rm Pl}^4} \ ,
\end{align}
and this requires
\begin{align}
 \frac{\zeta}{V_{,\phi}}
 = \frac{\beta}{\alpha} \gg 1 \ . \label{zetadV}
\end{align}
In this case, $\alpha$ becomes negligible. Hence
\begin{align}
 n_{\rm S} - 1= 2\gamma
 = 2M_{\rm Pl}^2 \frac{\zeta_{,\phi}}{V} \ ,\quad
 n_{\rm T} = 0 \ ,\quad
 r = 16\beta
 = 8M_{\rm Pl}^2\frac{\zeta^2}{V^2}\ . \label{obs_flat}
\end{align}
These equations means that if the Gauss-Bonnet correction 
dominates this equation of motion, the tensor-to-scalar ratio $r$
can take an observable value although $n_{\rm T}=0$ is realized.
Because of the consistency relation $r=-8n_{\rm T}$,
this scale invariant spectrum is impossible in a conventional
model.
Therefore, if scale invariant primordial gravitational waves
are detected, it implies existence of the Gauss-Bonnet term
in an inflationary theory and $\xi_{,\phi}$ is determined
by Eq.(\ref{xiphi}) with $\delta=1$:
\begin{align}
 M_{\rm Pl}\xi_{,\phi}=\frac{\sqrt{2}}{\pi^2}
 \frac{1}{\sqrt{r}{\cal P}_{\rm S}}
 = 1.86\times 10^8
 \left(\frac{{\cal P}_{\rm S}}{2.441\times 10^{-9}}\right)^{-1}
 \left(\frac{r}{0.1}\right)^{-1/2} \ ,
\end{align}
and using Eq.(\ref{V_o})
\begin{align}
 \frac{V\xi_{,\phi}}{M_{\rm Pl}^3}
 = \frac{3}{\sqrt{2}}\sqrt{r}
 = 0.671 \left(\frac{r}{0.1}\right)^{1/2} \ .
\end{align}
As in a blue case, $\xi_{,\phi}\sim 10^8/M_{\rm Pl}$
and $\xi_{,\phi}\sim M_{\rm Pl}^3/V$ must be satisfied in this flat
spectrum case, too.
Both too large and too small a value of 
$\xi_{,\phi}$ cannot lead to
a consistent scale invariant spectrum.
From Eqs.(\ref{zetadV}) and (\ref{obs_flat}),
the inequality for $r$ holds:
\begin{align}
 r \gg 8 M_{\rm Pl}^2 \frac{V_{,\phi}^2}{V^2} \ .
\end{align}
This shows that potential must be nearly flat for the 
tensor-to-scalar ratio $r$ to remain small.

Here, we check whether or not an observationally consistent
scale invariant model can be constructed.
In this subsection, we also use two types of examples,
namely a simple and analytic one and a realistic one.
Let us start with an analytic model:
\begin{align}
 V = V_0 
 \left| \tanh\left(\frac{\phi}{\phi_0} \right)\right|
 \ , \quad
 \xi_{,\phi}(\phi) = {\rm const.}
\end{align}
The e-folding number $N$ becomes
\begin{align}
 N = \frac{\phi_0^2}{2M_{\rm Pl}^2} \frac{1}{\theta}
 \ln(\theta x^2+ 1) \ ,
\end{align}
where $x$ and $\theta$ are defined by
\begin{align}
 x \equiv \sinh\left(\frac{\phi}{\phi_0}\right)
 \ , \quad
 \theta \equiv \frac{V_0\xi_{,\phi}\phi_0}{6M_{\rm Pl}^4}
 \ .
\end{align}
We use $x$ instead of $\phi$, and $x$ is
\begin{align}
 x^2 = \frac{1}{\theta}
 \left[
 \exp\left(\frac{2M_{\rm Pl}^2}{\phi_0^2} \theta N\right) - 1
 \right] \ .
\end{align}
The observables, namely the scalar spectral index $n_{\rm S}$, 
the tensor spectral index $n_{\rm T}$ and 
the tensor-to-scalar ratio $r$ become
\begin{align}
 n_{\rm S} - 1 &= \frac{M_{\rm Pl}^2}{\phi_0^2}
 \frac{1}{x^2+1}
 \left(-\frac{3}{x^2}-4+\theta\right) \\
 n_{\rm T} &= \frac{M_{\rm Pl}^2}{\phi_0^2} \frac{1}{x^2+1}
 \left(-\frac{1}{x^2}-\theta\right) \\
 r &= \frac{8M_{\rm Pl}^2}{\phi_0^2}
 \frac{1}{x^2+1} \left(\frac{1}{x}+\theta x\right)^2 \ .
\end{align}
In this model, we also variate $\theta$ and fix $\phi_0$.

Let us consider the limit case, in which
$\theta\ll \phi_0^2/2M_{\rm Pl}^2N$ is satisfied, 
and $x$ becomes
\begin{align}
 x^2 = \frac{2M_{\rm Pl}^2}{\phi_0^2} N \ .
\end{align}
Hence, observables are
\begin{align}
 n_{\rm S} - 1 = -\frac{1}{2N} \frac{4x^2+3}{x^2+1}
 \ , \quad
 n_{\rm T} = -\frac{1}{2N}\frac{1}{x^2+1}
 \ , \quad
 r = \frac{4}{N} \frac{1}{x^2+1} \ .
\end{align}
We can see that in this small $\theta$ limit, all models give
observationally consistent values.
Next, we investigate the large $\theta$ limit,
$\theta\gg \phi_0^2/2M_{\rm Pl}^2 N$.
The variable $x$ takes 
an exponentially large value in this limit. 
Hence, observables become
\begin{align}
 n_{\rm S} - 1= \frac{M_{\rm Pl}^2}{\phi_0^2} \frac{\theta}{x^2}
 \ , \quad
 n_{\rm T} = -\frac{M_{\rm Pl}^2}{\phi_0^2} \frac{\theta}{x^2}
 \ , \quad
 r = \frac{8M_{\rm Pl}^2}{\phi_0^2} \theta^2 \ .
\end{align}
These show that in this limit, $n_{\rm S}$ and $n_{\rm T}$
approach to flat one and $r$ becomes  large.
Remember that $x$ behaviors exponentially as a function of
$\theta$. 
If this large $\theta$ limit is realized in the WMAP consistent
region, such as $r<0.3$, the
result which has an observable amount of $r$ and flat 
$n_{\rm T}$ is achieved. 
Here, we consider $\theta=\phi_0^2/2M_{\rm Pl}^2N$ case, which
is the boundary case between large and small $\theta$.
In this case, $1+x^2\theta={\rm e}$ is realized.
The observables $n_{\rm T}$ and $r$ become
\begin{align}
 n_{\rm T} = - \frac{1}{2N} \frac{{\rm e}}{{\rm e}-1}
 \frac{\theta}{{\rm e}-1+\theta}
 = -0.0132  \frac{\theta}{{\rm e}-1+\theta}
 \ , \quad
 r = \frac{4}{N} \frac{{\rm e}^2}{{\rm e}-1}
 \frac{\theta}{{\rm e}-1+\theta}
 = 0.287  \frac{\theta}{{\rm e}-1+\theta} \ .
\end{align}
If $\theta=\phi_0^2/2M_{\rm Pl}^2N> {\rm e}-1$ is satisfied, 
$-n_{\rm T}>0.00659$ and
$r>0.143$ hold.
This $r$ might be too large to realize that large 
$\theta$ limit in $r<0.3$. Therefore, to achieve desirable
results, $\theta\ll{\rm e}-1$ at this
$\theta=\phi_0^2/2M_{\rm Pl}^2N$ case might be necessary. 
In other words,
\begin{align}
 \phi_0 \ll \sqrt{2N({\rm e}-1)} M_{\rm Pl}
 = 14.4 M_{\rm Pl} \ .
\end{align}
The physical meaning of this condition
can be easily understood: Large field model requires
large $\dot{\phi}$ and inconsistently 
large $r(>0.3)$, to reach a flat potential
region within $N=60$ e-fold. 
This interpretation is essentially the same
as that in the condition for blue spectrum.
The difference is that small field model can 
also achieve large $r$ in this case, 
because {\it small} doesn't means the
smallness of the variation of the field value of the 
inflaton $\phi$.
It means the smallness of the field value between the 
vacuum and a flat
region of potential. Hence, this is consistent with
the Lyth bound.
If this condition is satisfied, sufficiently large $\theta$,
namely $\theta\gg\phi_0^2/2M_{\rm Pl}^2N$, in other words
\begin{align}
 \frac{V_0 \xi_{,\phi}}{M_{\rm Pl}^3} \gg 
 \frac{3}{N}\frac{\phi_0}{M_{\rm Pl}}
 \ ,
\end{align} 
leads to a consistent
flat spectrum.

\begin{figure}[htbp]
 \centering
 \includegraphics[scale=0.6]{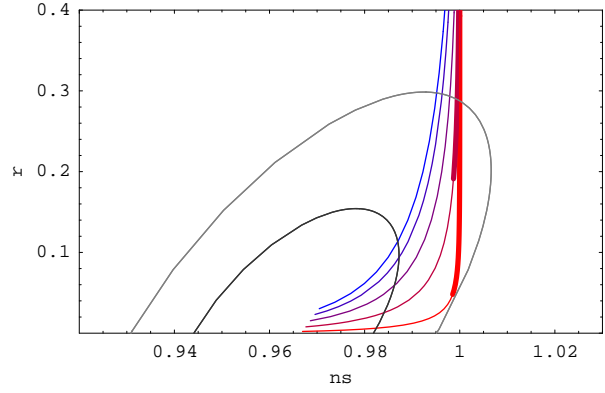}
 \caption{The trajectories, in which $\theta$ is varying,
 of analytic flat models are
 shown on a $n_{\rm S}-r$ plane.
 Two contours roughly corresponds to the 
 68\% and 95\% confidence level of the WMAP 7-year 
 result.
 We draw five lines, which are corresponding to
 $\phi_0=2$ (red, bottom) to $10M_{\rm Pl}$ (blue, top) 
 with an even interval. Thick lines denote the regions,
 which realize $r/|8n_{\rm T}| > 100$.
 Small $\phi_0$ models, namely $\phi_0=2M_{\rm Pl}$
 and $\phi_0=4M_{\rm Pl}$ models, realize observationally 
 consistent flat spectra.
 }
 \label{trjtan}
\end{figure}

We show  
$\theta$-trajectories of solutions on a $n_{\rm S}-r$ plane
in FIG.\ref{trjtan}.
We plot the 68\% and 95\% C.L.
contours from the WMAP 7-year result. 
We choose five $\phi_0$ values,
$2,\ 4,\ 6,\ 8$ and $10M_{\rm Pl}$.
Thick lines denote the flat region, in which $r/8|n_{\rm T}|>100$
is realized.
Note that $r=-8n_{\rm T}$ holds in conventional single
field inflation.
Therefore large $r/8|n_{\rm T}|$ value implies
a flat spectrum.
We can see that small $\phi_0$ models, namely 
$\phi_0=2M_{\rm Pl}$ and $\phi_0=4M_{\rm Pl}$ models,
result in  WMAP-consistent flat spectra.
This is consistent with the criterion at  the
above calculation, namely $\phi_0\ll 14.4M_{\rm Pl}$.
As mentioned above, $n_{\rm S}$ also becomes flat, in 
a flat tensor region.
Note that the smaller field model, 
namely $\phi_0=2M_{\rm Pl}$ model, has larger consistent
scale invariant region.
Of course, $\xi_{,\phi}$ can be reconstructed from 
Eq.(\ref{xiphi}), and $\xi_{,\phi}\sim10^8/M_{\rm Pl}$
must hold in a flat region.

Here, we consider
a realistic example of a flat spectrum model.
Nearly flat potential results in
a Gauss-Bonnet dominating situation. Therefore we take
the following potential and coupling function:
\begin{align}
 V(\phi) =  
 \rho 10^{-10} M_{\rm Pl}^4
 \left[
 1 - \exp\left(-\frac{\phi^2}{\phi_0^2}\right)
 \right]
 \ , \quad
 \xi(\phi)
 = s\frac{b}{\rho} \exp(sa \phi/M_{\rm Pl}) \ ,
\end{align}
where $s=\pm1$, $\rho$ will be fixed by 
${\cal P}_{\rm S} = 2.441 \times 10^{-9}$,
and $a, b>0$. 
In the $\phi\rightarrow\infty$ limit, this potential becomes
extremely flat.
We use the same exponential coupling in this example
as in previous subsection.

\begin{figure}[htbp]
 \centering
 \begin{minipage}{0.49\hsize}
  \centering
  \includegraphics[scale=0.6]{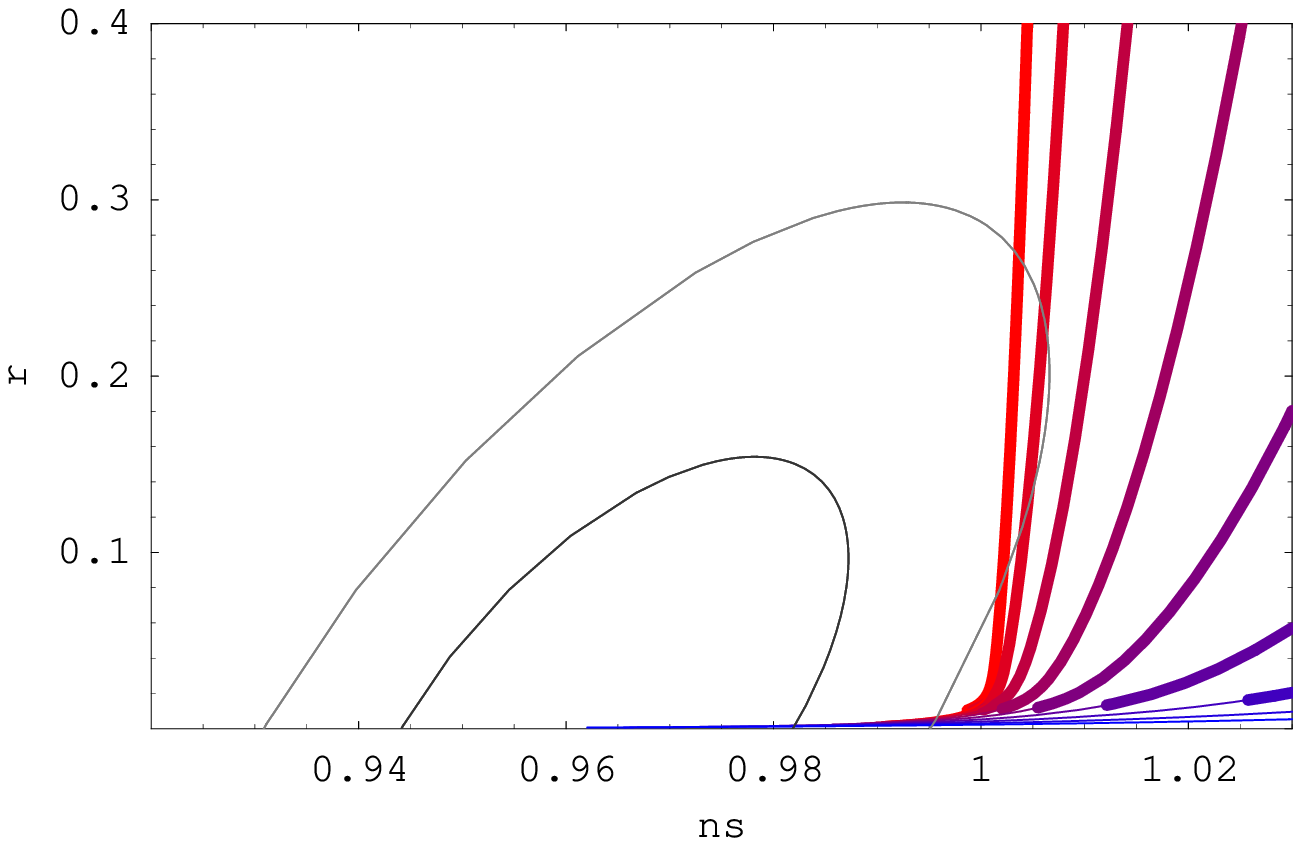}
  $\phi_0=2 M_{\rm Pl},\ s=+1$
  \vspace*{0.3cm}
 \end{minipage}
 \begin{minipage}{0.49\hsize}
  \centering
  \includegraphics[scale=0.6]{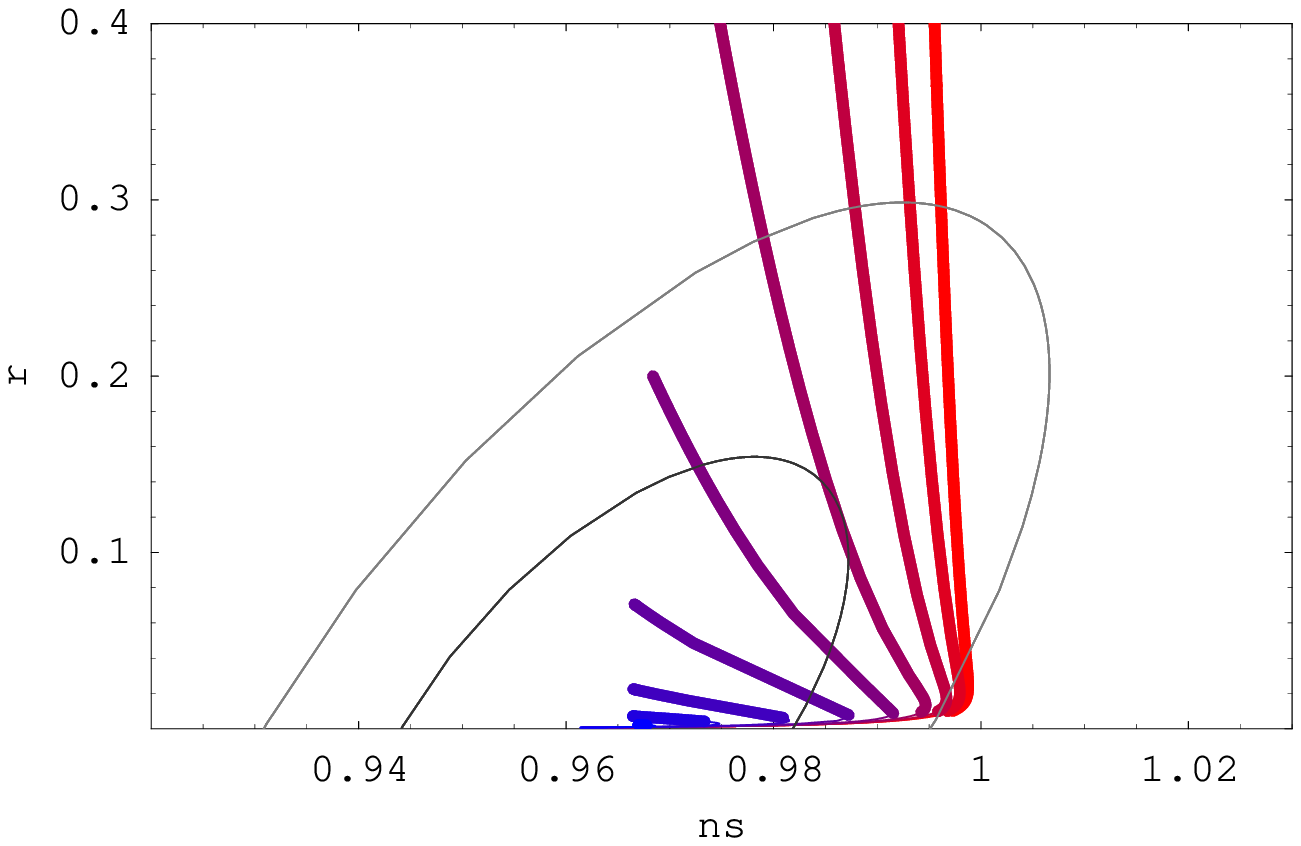}
  $\phi_0=2 M_{\rm Pl},\ s=-1$
  \vspace*{0.3cm}
 \end{minipage}
 \begin{minipage}{0.49\hsize}
  \centering
  \includegraphics[scale=0.6]{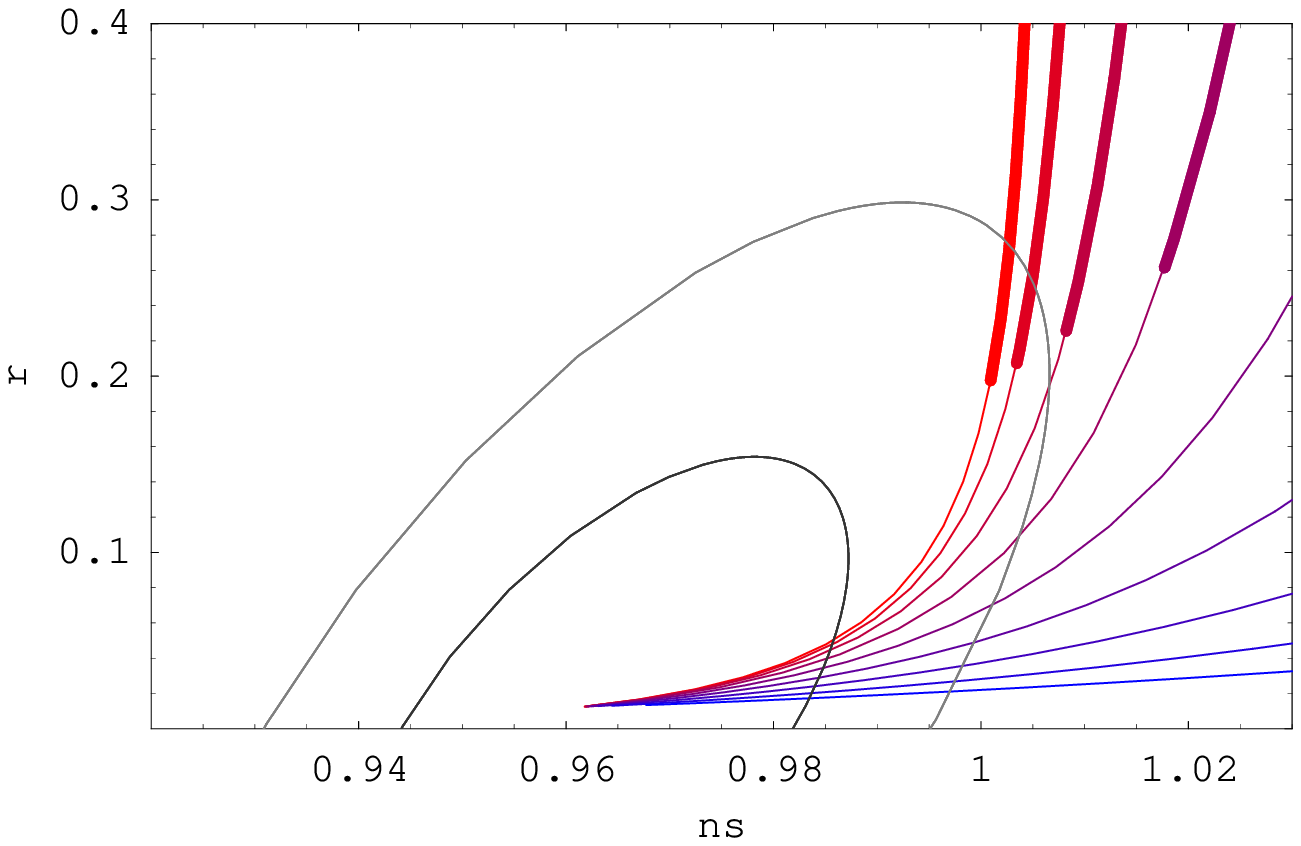}
  $\phi_0=6 M_{\rm Pl},\ s=+1$
  \vspace*{0.3cm}
 \end{minipage}
 \begin{minipage}{0.49\hsize}
  \centering
  \includegraphics[scale=0.6]{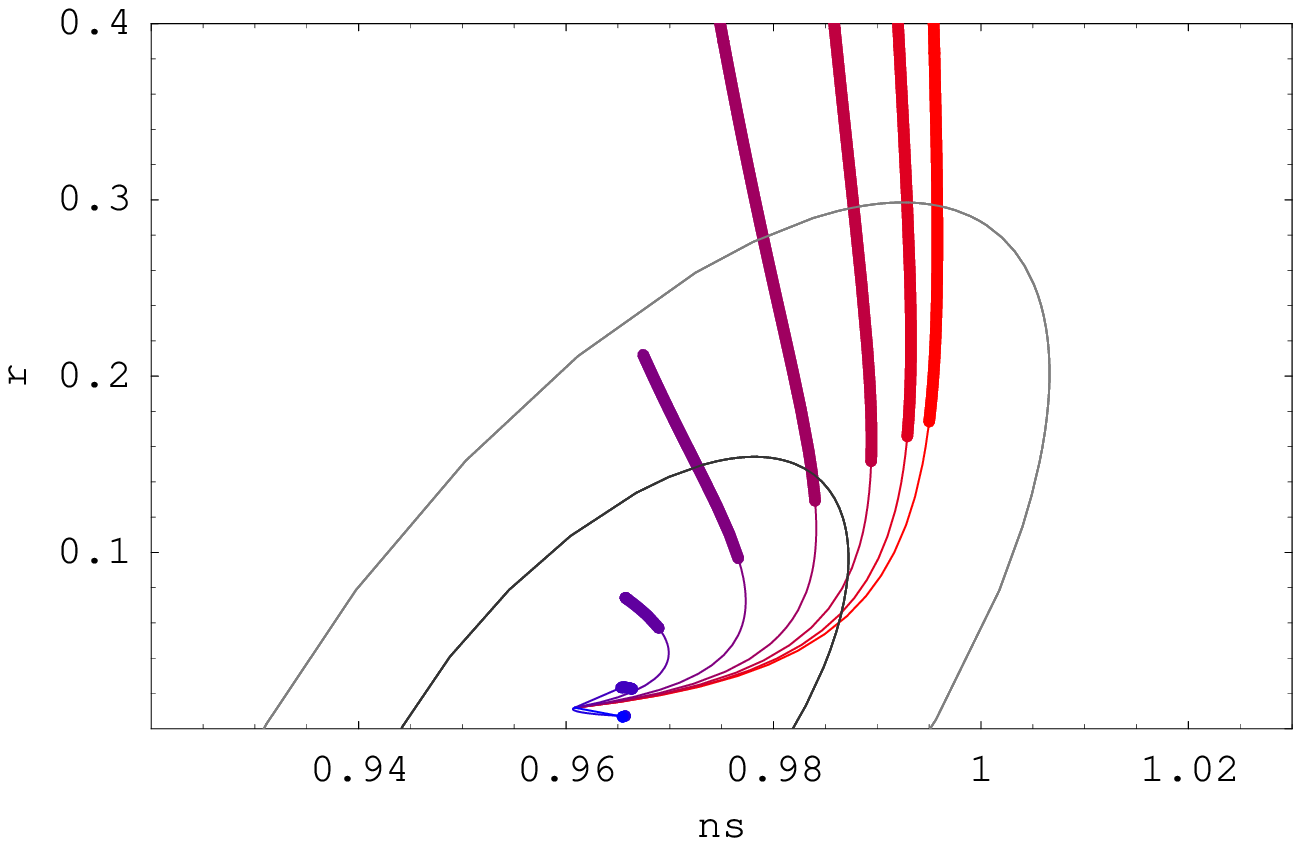}
  $\phi_0=6 M_{\rm Pl},\ s=-1$
  \vspace*{0.3cm}
 \end{minipage}
 \begin{minipage}{0.49\hsize}
  \centering
  \includegraphics[scale=0.6]{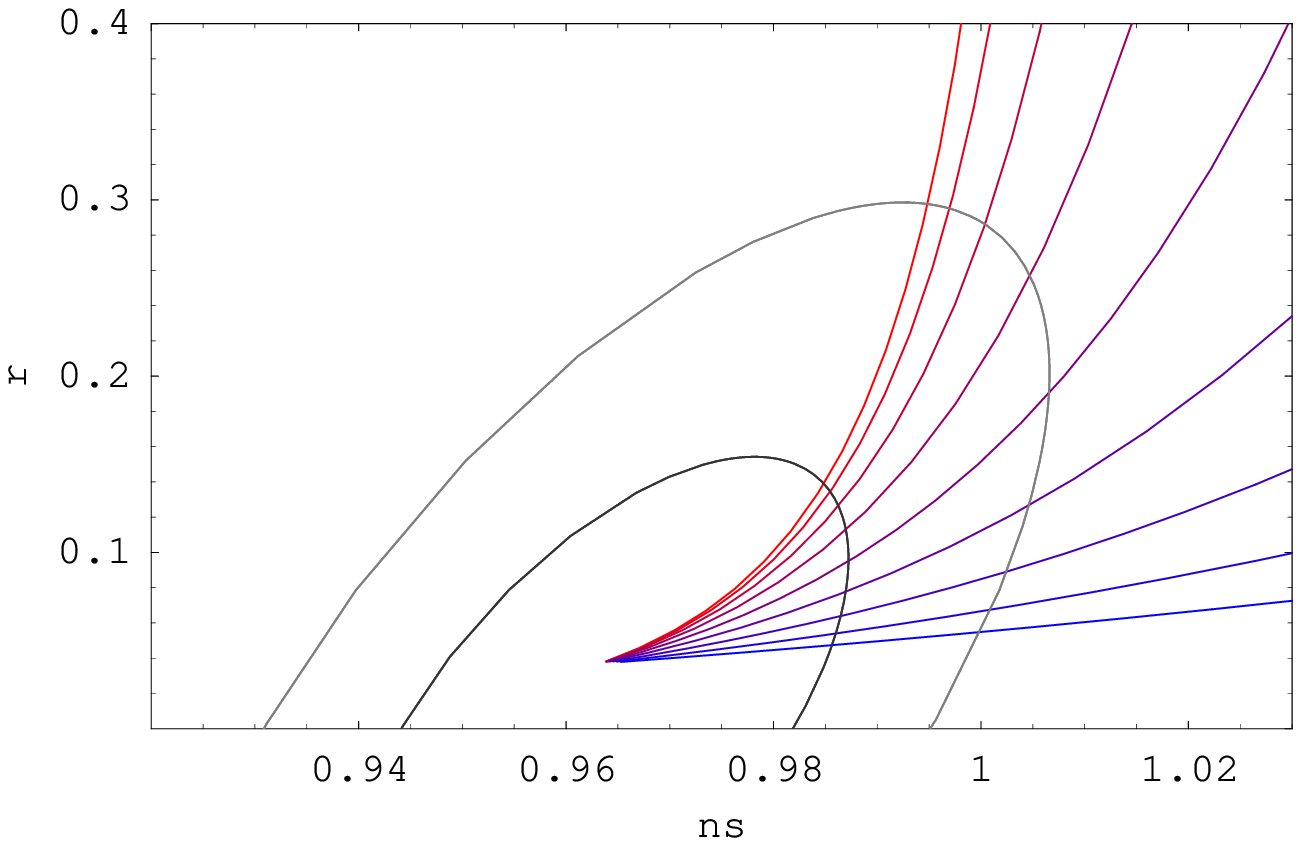}
  $\phi_0=10 M_{\rm Pl},\ s=+1$
  \vspace*{0.3cm}
 \end{minipage}
 \begin{minipage}{0.49\hsize}
  \centering
  \includegraphics[scale=0.6]{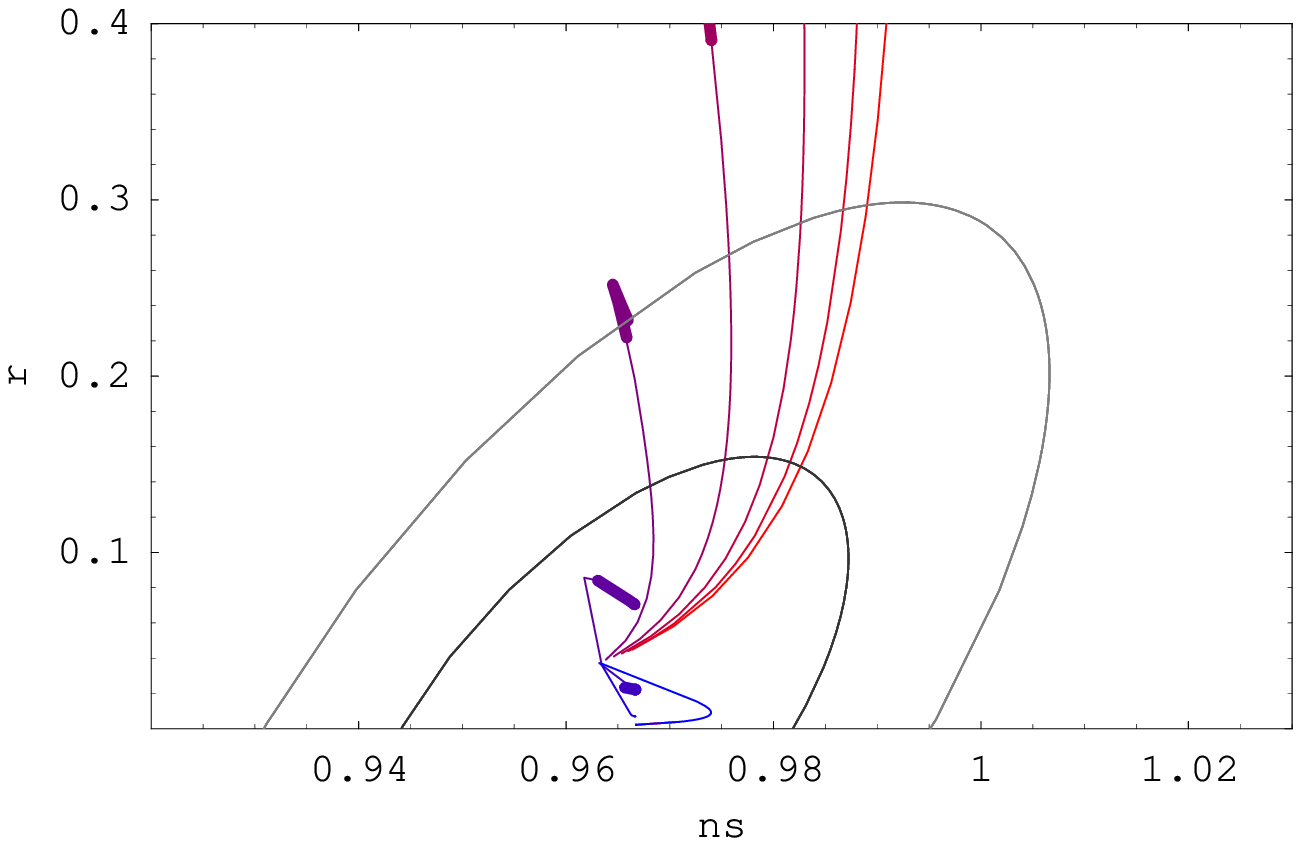}
  $\phi_0=10 M_{\rm Pl},\ s=-1$
  \vspace*{0.3cm}
 \end{minipage}
 \caption{The trajectories, in which $b$ is varying,
 of some flat models 
 is shown on a $n_{\rm S}-r$ plane.
 In the $b\rightarrow0$ limit, all trajectories converges
 to one point.
 Two contours roughly corresponds to
 the 68\% and 95\% confidence level of the WMAP 7-year 
 result.
 We draw nine lines, which are corresponding to
 $a=0.01$ (red, left in left panels and right in right panels)
 to $1$ (blue, right in left panels and left in right panels)
 with an even logarithmic
 interval. In this case, tensor spectrum
 is always red.
 Thick lines denote the regions of flat spectra, 
 $r/8|n_{\rm T}|>100$.
 This figure shows that small $\phi_0$ leads to a flat spectrum.
 Note that 
 in $s=-1$ models, trajectories approach to some points,
 in large $b$ limit.
 }
 \label{trjflat}
\end{figure}

We show $b$-trajectories
on a $n_{\rm S}-r$ plane
in FIG.\ref{trjflat}.
We plot the WMAP 68\% and 95\% C.L. contours.
We use three values of $\phi_0$, $2,\ 6$ and
$10M_{\rm Pl}$, and the same set of
$a$ in this case as in the realistic blue spectrum case.
Thick lines denote
the regions $r/8|n_{\rm T}|>100$.
As predicted from 
the previous analytic models, smaller $\phi_0$ models
has larger regions of flat spectra.
Note that
$s=-1$ models have larger consistent flat region
than that of $s=-1$ models.
An order unity value of $a$ leads to extremely small
$r$ in both $s=\pm1$ models, and
considering reheating, large $a$ in $s=-1$ models
is not desirable.
Therefore also in this case, $a\ll1$ is appropriate.
Eq.(\ref{xiphi}) is valid to get the value of 
$\xi_{,\phi}$.

In $s=-1$ models, trajectories approach to some
points for each $a$ in large $b$ limit, in FIG.\ref{trjflat}.
This is understood from Eq.(\ref{e-fold_0}).
In the large $b$ limit of $s=-1$ models, 
only a large $\phi$ region contribute to the integration 
for an e-folding number.
Because $V_{,\phi}/V\simeq0$ in this region,
we can omit $V_{,\phi}/V$ term: 
\begin{align}
 N = \frac{6 M_{\rm Pl}^2}{V_0}
 \int^\phi_0 \frac{{\rm d}\phi}{\xi_{,\phi}}
 = \frac{6M_{\rm Pl}^3}{aV_0\xi_{,\phi}} \ .
\end{align}
This shows that, in this large $b$ limit, $\xi_{,\phi}$
approaches to $\xi_{,\phi}=6M_{\rm Pl}^3/aNV_0$.
This also fixes the value of $\xi_{,\phi\phi}$ and all slow-roll 
parameters. The observables $n_{\rm S}$ and $r$ become
\begin{align}
 n_{\rm S} - 1 = \frac{V_0\xi_{,\phi\phi}}{3M_{\rm Pl}^2}
 = -\frac{2}{N}
 \ ,\quad
 r = \frac{2 V_0^2\xi_{,\phi}^2}{9 M_{\rm Pl}^6}
 = \frac{8}{a^2N^2}
 \ .
\end{align}
Hence, trajectories approach to some points in large $b$.
This shows that $a\simeq1$ model leads to very small $r$ 
because $r$ approaches to $r=8/N^2=0.00222$ at $a=1$. 

From above results, we can say
that an observationally consistent flat spectrum can be 
obtained.
Hence, confirming existence of the
Gauss-Bonnet term is possible, with observing
a scale invariant tensor mode.
In the cases we considered, 
$\phi_0\ll 10M_{\rm Pl}$ and nearly constant
$\xi_{,\phi}$ might lead to a consistent flat spectrum.
Note that $\phi_0$ denotes the difference between
the filed value of a flat region and the vacuum of 
potential.
We can expect these conditions are
also satisfied in many other cases.
To realize a flat spectrum, 
$\xi_{,\phi}\sim 10^8/M_{\rm Pl}$ is necessary in any model.

\afterpage{\clearpage}
\subsection{Inflation with steep potential}
As mentioned in the end of section \ref{sec_slow-roll},
if we take $\xi=6M_{\rm Pl}^4/V$, potential is cancelled out
by the Gauss-Bonnet effective potential.
Considering this fact, let us take $\xi$ as follows:
\begin{align}
 \xi = (1-\kappa) \frac{6M_{\rm Pl}^4}{V} \ ,
 \label{xi_cance}
\end{align}
where $\kappa$ is a constant and we assume
$|\kappa|\ll1$. 
The function $\zeta$ becomes
\begin{align}
 \zeta = V_{,\phi}
 - (1-\kappa) V_{,\phi}
 = \kappa V_{,\phi} \ .
\end{align}
Eq.(\ref{dphizeta})
becomes
\begin{align}
 \frac{\dot{\phi}}{M_{\rm Pl}H}
 = - \kappa M_{\rm Pl} \frac{V_{,\phi}}{V} \ .
\end{align}
Hence if $\kappa<0$ is realized, a scalar field always
rolls up potential toward a hill-top.
This is not a desirable situation. Therefore we take
$0<\kappa\ll1$.
Slow-roll parameters becomes
\begin{align}
 \alpha = \kappa \frac{M_{\rm Pl}^2}{2}
 \frac{V_{,\phi}^2}{V^2} \ , \quad
 \beta = \kappa^2 \frac{M_{\rm Pl}^2}{2}
 \frac{V_{,\phi}^2}{V^2} \simeq 0 \ , \quad
 \gamma = \kappa M_{\rm Pl}^2
 \frac{V_{,\phi\phi}}{V} \ ,
\end{align}
where we used $\kappa\ll1$.
In this case, $\beta$ becomes negligible.
Note if the Gauss-Bonnet term is dominating, $\alpha$ becomes
negligible. 
In this case, the observables $n_{\rm S}$, $n_{\rm T}$
and $r$ are
\begin{align}
 n_{\rm S} - 1 = -6\alpha + 2\gamma \ , \quad
 n_{\rm T} = -2\alpha \ , \quad
 r = 0 \ .
\end{align}
Because the tensor-to-scalar ratio $r$ is zero, the value
of the spectral
index $n_{\rm T}$ is meaningless.
Therefore observationally speaking, 
this situation is not interesting.
However, because $\kappa\ll 1$, three slow-roll conditions are
automatically satisfied. 
Hence, even if we use extremely steep
potential, inflation can be achieved.

\section{Conclusion}
In this paper, we studied slow-roll inflation with
the Gauss-Bonnet and Chern-Simons corrections.
We defined three slow-roll parameters, while we used
five parameters in the previous paper\cite{satoh08b}.
We derived  
expressions for 
the scalar spectral index $n_{\rm S}$, 
the tensor spectral index
$n_{\rm T}$, 
the tensor-to-scalar ratio $r$ and 
the circular polarization ratio $\Pi$
by using these parameters.
We showed that in our model, the consistency relation
$r=-8n_{\rm T}$ is not automatically satisfied.
If this violation is observationally confirmed, 
we can determine the derivative 
of the Gauss-Bonnet coupling function $\xi_{,\phi}$
at the CMB scale.
In our model, 
even both blue and scale invariant spectra of 
gravitational waves can be realized.
Because blue and scale invariant mean $n_{\rm T}>0$
and $|8n_{\rm T}|/r \ll 1$, respectively,
these cases violate this consistency relation strongly.
Therefore these are the key for confirming our model in
future observations.
We showed that 
if either a blue spectrum or a scale invariant one
is observed, it supports existence of 
the Gauss-Bonnet coupling function of
 $\xi_{,\phi}\sim10^8/M_{\rm Pl}$ at the CMB scale.
We checked whether or not
 these blue and scale invariant spectra
are consistent with current observations.
For this purpose, we used concrete examples.
We showed that new-inflation-type potential with
$10M_{\rm Pl}$ symmetry breaking scale and 
potential with flat region in $\phi\gtrsim 10M_{\rm Pl}$
might result in observationally consistent
blue and scale invariant spectra,
respectively. 
An almost linear form of the Gauss-Bonnet coupling 
function is appropriate in both cases.
These tell us that the detection of a blue or scale invariant
spectrum of tensor modes can be actually expected with
future observations. 
We also showed that if circular polarization of gravitational
waves is detected, the derivative of the Chern-Simons coupling 
$\omega_{,\phi}$ must be on the order of $10^8/M_{\rm Pl}$. 
Thus, we can say that existence
of gravitational higher coupling terms in a inflationary 
model will be confirmed, or at least
constrained, with future 
experiments.

These higher curvature terms are expected to appear in
non-Gaussian part of perturbations. 
Therefore, calculating non-Gaussianity in our
model is important future work.

\vskip 2cm
We would like to
thank Jiro Soda for his very useful comments.


\begin{thebibliography}{99}
 \bibitem{komatsu10}
	 E.~Komatsu {\it et al.},
	 arXiv:1001.4538[astro-ph.CO].

\bibitem{Antoniadis:1993jc}
  I.~Antoniadis, J.~Rizos and K.~Tamvakis,
  Nucl.\ Phys.\  B {\bf 415}, 497 (1994)
  [arXiv:hep-th/9305025].

%
\bibitem{Kawai:1998bn}
  S.~Kawai and J.~Soda,
  Phys.\ Rev.\  D {\bf 59}, 063506 (1999)
  [arXiv:gr-qc/9807060].


 

%
\bibitem{Kawai:1999xn}
  S.~Kawai and J.~Soda,
  arXiv:gr-qc/9906046.

%
\bibitem{Kawai:1999pw}
  S.~Kawai and J.~Soda,
  Phys.\ Lett.\  B {\bf 460}, 41 (1999)
  [arXiv:gr-qc/9903017].
  
\bibitem{Hwang:1999gf}
  J.~c.~Hwang and H.~Noh,
  Phys.\ Rev.\  D {\bf 61}, 043511 (2000)
  [arXiv:astro-ph/9909480].
  
%
\bibitem{Cartier:2001is}
  C.~Cartier, J.~c.~Hwang and E.~J.~Copeland,
  Phys.\ Rev.\  D {\bf 64}, 103504 (2001)
  [arXiv:astro-ph/0106197].
 
%
\bibitem{Kawai:1998ab}
  S.~Kawai, M.~a.~Sakagami and J.~Soda,
  Phys.\ Lett.\  B {\bf 437}, 284 (1998)
  [arXiv:gr-qc/9802033].

%
\bibitem{Soda:1998tr}
  J.~Soda, M.~a.~Sakagami and S.~Kawai,
  arXiv:gr-qc/9807056.
  
%
\bibitem{Kawai:1997mf}
  S.~Kawai, M.~a.~Sakagami and J.~Soda,
  arXiv:gr-qc/9901065.

 %
\bibitem{Gasperini:1997up}
  M.~Gasperini,
  Phys.\ Rev.\  D {\bf 56}, 4815 (1997)
  [arXiv:gr-qc/9704045].
  
%
\bibitem{Cartier:2001gc}
  C.~Cartier, E.~J.~Copeland and M.~Gasperini,
  Nucl.\ Phys.\  B {\bf 607}, 406 (2001)
  [arXiv:gr-qc/0101019].

\bibitem{Leith:2007bu}
  B.~M.~Leith and I.~P.~Neupane,
  JCAP {\bf 0705}, 019 (2007)
  [arXiv:hep-th/0702002].
%
\bibitem{Guo07}
  Z.~K.~Guo, N.~Ohta and S.~Tsujikawa,
  Phys.\ Rev.\ D {\bf 75}, 023520 (2007),
  [arXiv:hep-th/0610336].  

\bibitem{Vazquez:2008wb}
 S.~E.~Vazquez,
 Phys.\ Rev.\ D {\bf 79}, 043520 (2009), 
 arXiv:0806.0603 [hep-th].

  
\bibitem{Nojiri:2005vv}
  S.~Nojiri, S.~D.~Odintsov and M.~Sasaki,
  Phys.\ Rev.\  D {\bf 71}, 123509 (2005)
  [arXiv:hep-th/0504052].
%
\bibitem{neupane06a}
  I.~P.~Neupane and B.~M.~N.~Carter,
  JCAP {\bf 0606}, 004 (2006)
  [hep-th/0512262].

 \bibitem{koivisto07a}
	 K.~T.~Koivisto and D.~F.~Mota,
	 Phys.\ Lett.\ B {\bf 644}, 104 (2007)
	 [arXiv:astro-ph/0606078].

 \bibitem{koivisto07b}
	 K.~T.~Koivisto and D.~F.~Mota,
	 Phys.\ Rev.\ D {\bf 75}, 023518 (2007)
	 [arXiv:hep-ph/0609115].

\bibitem{Neupane08}
	I.~P.~Neupane, arXiv:0711.3234[hep-th].
\bibitem{neupane06b}
  I.~P.~Neupane,
  Class.\ Quant.\ Grav. {\bf 23}, 7493 (2006) 
  [hep-th/0602097]. 

 \bibitem{defelice09a}
	 A.~D.~Felice and S.~Tsujikawa,
	 Phys.\ Lett.\ B {\bf 675}, 1 (2009),
	 arXiv:0810.5712[hep-th].

 \bibitem{defelice09b}
	 A.~D.~Felice and S.~Tsujikawa,
	 Phys.\ Rev.\ D {\bf 80}, 063516 (2009),
	 arXiv:0907.1830[hep-th].
	 
 \bibitem{defelice10a}
	 A.~D.~Felice, D.~F.~Mota and S.~Tsujikawa,
	 Phys.\ Rev.\ D {\bf 81}, 023532 (2010),
	 arXiv:0911.1811[gr-qc].

 \bibitem{defelice10b}
	 A.~D.~Felice and S.~Tsujikawa,
	 Living Rev.\ Rel.\ {\bf 13}, 3 (2010),
	 arXiv:1002.4928[gr-qc].



\bibitem{Lue:1998mq}
  A.~Lue, L.~M.~Wang and M.~Kamionkowski,
  Phys.\ Rev.\ Lett.\  {\bf 83}, 1506 (1999)
  [arXiv:astro-ph/9812088].

%
\bibitem{Choi:1999zy}
  K.~Choi, J.~c.~Hwang and K.~W.~Hwang,
  Phys.\ Rev.\  D {\bf 61}, 084026 (2000)
  [arXiv:hep-ph/9907244].

%
\bibitem{Alexander:2004wk}
  S.~Alexander and J.~Martin,
  Phys.\ Rev.\  D {\bf 71}, 063526 (2005)
  [arXiv:hep-th/0410230].

%
\bibitem{Lyth:2005jf}
  D.~H.~Lyth, C.~Quimbay and Y.~Rodriguez,
  JHEP {\bf 0503}, 016 (2005)
  [arXiv:hep-th/0501153].

\bibitem{satoh08a}
  M.~Satoh, S.~Kanno and J.~Soda,
 Phys.\ Rev.\ D\ {\bf 77}, 023526 (2008), 
  arXiv:0706.3585[astro-ph].

%
\bibitem{Saito:2007kt}
  S.~Saito, K.~Ichiki and A.~Taruya,
  JCAP {\bf 0709}, 002 (2007),  
  arXiv:0705.3701 [astro-ph].
  
%
\bibitem{Seto:2006hf}
  N.~Seto,
  Phys.\ Rev.\ Lett.\  {\bf 97}, 151101 (2006)
  [arXiv:astro-ph/0609504].
  
%
\bibitem{taruya}
N.~Seto and A.~Taruya,
Phys.\ Rev.\ Lett.\ {\bf 99}, 121101 (2007),
arXiv:0707.0535 [astro-ph].
%

 \bibitem{weinberg08}
	 S.~Weinberg,
	 Phys.\ Rev.\ D {\bf 77}, 123541 (2008),
	 arXiv:0804.4291[hep-th].
 \bibitem{satoh08b}
	 M.~Satoh and J.~Soda,
	 JCAP {\bf 0809}, 019 (2008),
	 arXiv:0806.4594[astro-ph].
 \bibitem{guo10}
	 Z.~K.~Guo and D.~J.~Schwarz,
	 Phys.\ Rev.\ D {\bf 81}, 123520 (2010),
	 arXiv:1001.1897[hep-th].
 \bibitem{mukhanov06}
	 V.~F.~Mukhanov and A.~Vikman,
	 JCAP {\bf 0602}, 004 (2006) 
	 [arXiv:astro-ph/0512066].
 \bibitem{lyth97}
	 D.~H.~Lyth,
	 Phys.\ Rev.\ Lett.\ {\bf 78}, 1861 (1997) 
	 [arXiv:hep-ph/9606387].
\end{thebibliography}
\end{document}